\documentclass[showpacs,twocolumn,superscriptaddress,prl,preprintnumbers,nofootinbib]{revtex4}
\usepackage{amsmath,amssymb,bm,graphicx,color,gensymb}

\begin{document}
\title{Field-induced decays in $XXZ$ triangular-lattice antiferromagnets}
\author{P. A. Maksimov}
\affiliation{Department of Physics and Astronomy, University of California, Irvine, California 92697, USA}
\author{M. E. Zhitomirsky}
\affiliation{CEA, INAC-PHELIQS, F-38000, Grenoble, France}
\author{A. L. Chernyshev}
\affiliation{Department of Physics and Astronomy, University of California, Irvine, California 92697, USA}

\date{\today}
\begin{abstract}
We investigate field-induced transformations in the dynamical response of the $XXZ$ model on the triangular lattice
that are associated with the anharmonic magnon coupling and decay phenomena. A set of concrete theoretical predictions
is made for a close physical realization of the spin-$\frac12$ $XXZ$ model, Ba$_3$CoSb$_2$O$_9$. 
We demonstrate that dramatic modifications in magnon spectrum must occur in low 
out-of-plane fields that are easily achievable for this material. 
The hallmark of the effect is a coexistence of the clearly distinct well-defined magnon excitations with   
significantly broadened ones in different regions of the ${\bf k}\!-\!\omega$ space. 
The field-induced decays are generic for this class of models and become more prominent at larger anisotropies 
and in higher fields.
\end{abstract}
\maketitle
Triangular-lattice antiferromagnets (TLAFs) are central to the field of frustrated magnetism as 
representatives of one of the basic models epitomizing the effect of spin frustration \cite{wannier,spinliquid,Huse88,lee84}.
They have attracted significant experimental and theoretical interest 
\cite{Collins97,Coldea01,Svistov03,Nakatsuji05,Olariu06,%
Oguchi83,Jolicoeur89,Miyake92,Chubukov94,Leung93,capriotti99,Singh91,zheng06,white07,ZhuWhite,%
starykh10,Lhuillier94} 
as a potential source of spin-liquid  and of a wide variety of intriguing ordered ground 
states, see Ref.~\cite{starykh_review}. Their spectral properties have recently emerged as a subject of intense research
that has consistently uncovered broad, continuum-like spectral features \cite{Coldea01,expBaCoSbO2016,JeGeun}, which 
are interpreted as an evidence of fractionalized excitations \cite{Coldea01,starykh10,oitmaa2015} 
or of the phenomenon of magnon decay \cite{RMP,starykh06,triangle,triSqw}.

\begin{figure}[b]
\includegraphics[width=0.99\linewidth]{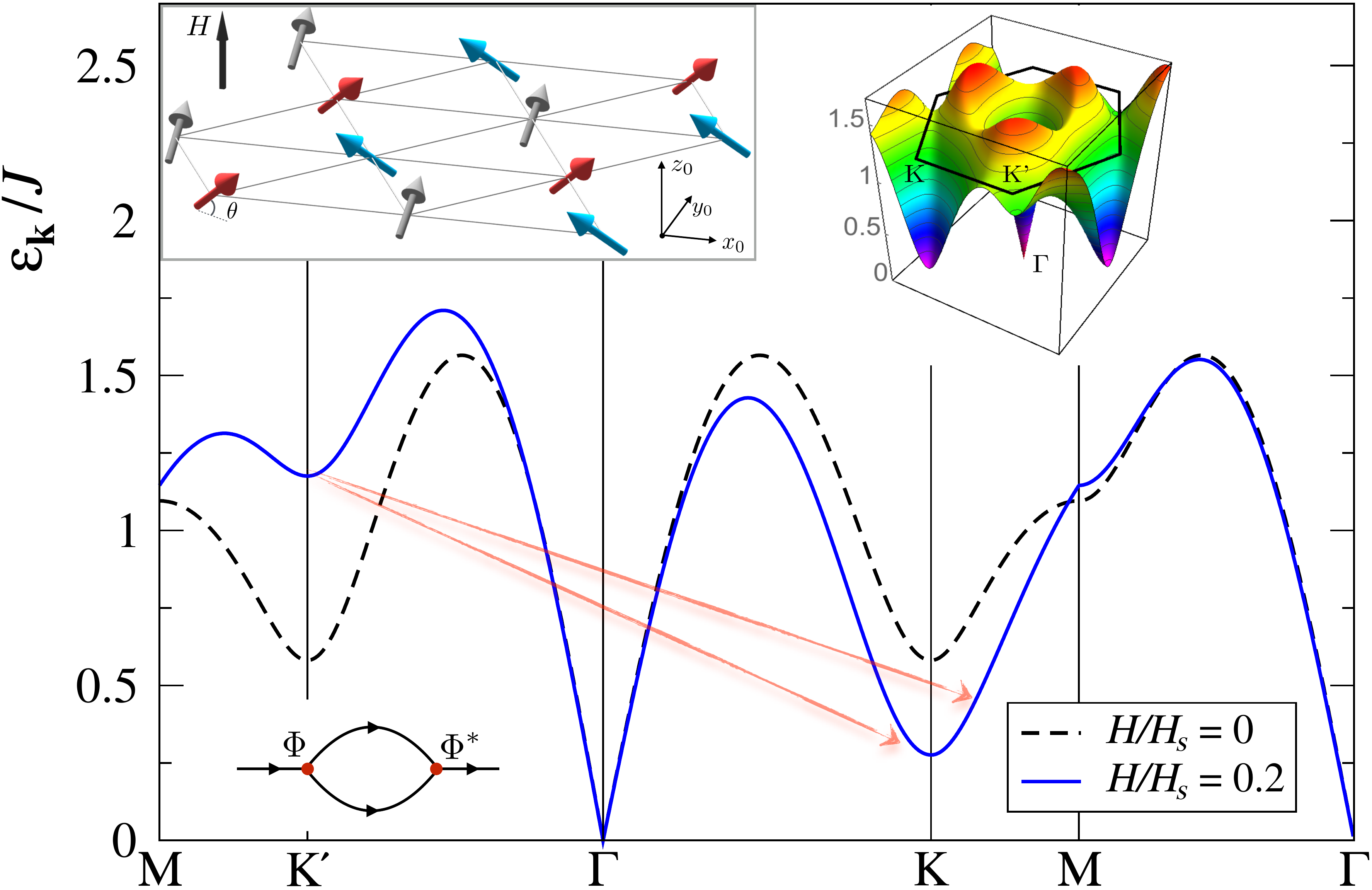}
\caption{Linear spin wave energy $\varepsilon_{\bf k}$ of model (\ref{model}) 
for $\Delta=0.9$, $S=1/2$ and the fields $H=0$ and $H=0.2H_s$. 
Arrows show schematics of the decay. Insets: umbrella structure in a field, 3D plot of $\varepsilon_{\bf k}$ 
for $H=0.2H_s$, and decay self-energy diagram.}
\label{fig_linearspectrum}
\end{figure}

In this work, we outline a theoretical proposal for a dramatic transformation of the spin-excitation spectrum
of the $XXZ$ triangular-lattice antiferromagnet in external out-of-plane field. This consideration pertains 
in particular to Ba$_3$CoSb$_2$O$_9$, one of the close physical realizations of the model that has recently been 
studied by a variety of experimental techniques 
\cite{expBaCoSbO2004,expBaCoSbO2012_1,expBaCoSbO2012_2,expBaCoSbO2013,expBaCoSbO2015}. 
The key finding of our work is that a modest out-of-plane field  results in a strong damping of the high-energy
magnons, affecting a significant part of the ${\bf k}$-space. This is different from a similar prediction of the 
field-induced decays in the square- and honeycomb-lattice AFs where strong spectrum transformations require
large fields \cite{99,Mourigal10,Fuhrman12,ushoney16}. In the present case, 
because the staggered chirality of the field-induced umbrella spin structure 
breaks inversion symmetry, the resultant ${\bf k}\leftrightarrow -{\bf k}$ asymmetry of the magnon spectrum 
opens up a channel for decays of the high-energy magnons in a broad vicinity of the K$^\prime$ corners 
of the Brillouin zone into the two-magnon continuum of the roton-like magnons at the $K$-points, see 
Fig.~\ref{fig_linearspectrum}.

We note that the recent neutron-scattering work \cite{expBaCoSbO2016} asserts the existence of an 
intrinsic broadening in parts of the Ba$_3$CoSb$_2$O$_9$ spectrum
even in zero field. While scatterings due to finite-temperature magnon population 
or strong effects of disorder in the non-collinear spin structures
\cite{Wolfram} cannot be ruled out as sources of damping observed in Ref.~\cite{expBaCoSbO2016}, 
we would like to point out that the phenomena discussed in this work 
are substantially more dramatic and should be free from such uncertainties.

\begin{figure}
\includegraphics[width=0.99\linewidth]{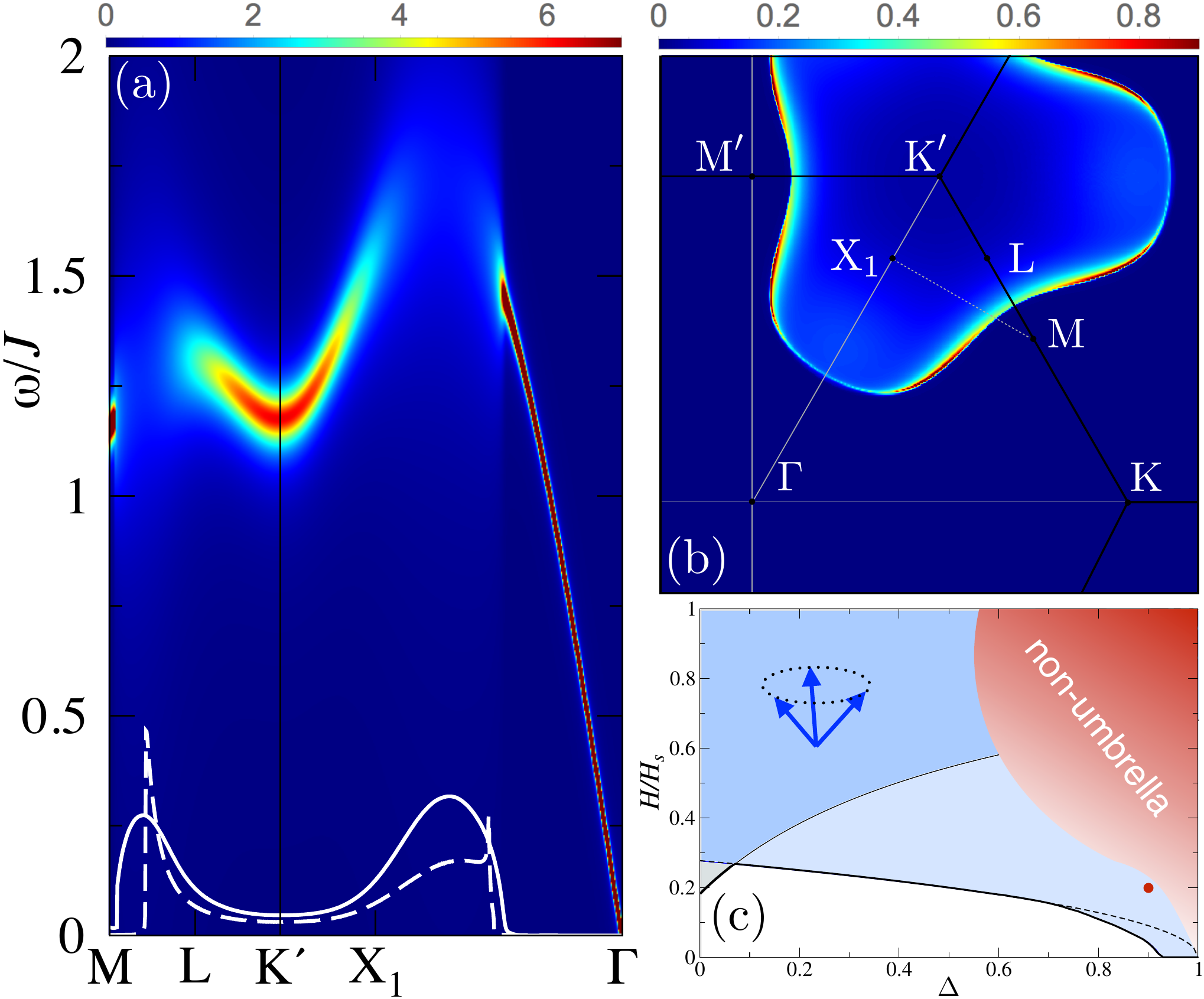}
\caption{(a) The intensity plot of the spectral function along MK$' \Gamma$ path 
with $\Gamma_{\bf k}$ from the self-consistent iDE for $\Delta\!=\!0.9$ and $H\!=\!0.2H_s$. 
Dashed and solid lines are $\Gamma_{\bf k}$ 
in the Born approximation (\ref{eq_gammak}) and the iDE solution. 
(b) The 2D intensity plot of $\Gamma_{\bf k}$ from \eqref{eq_gammak}. 
(c) The $H\!-\!\Delta$ diagram of the decay thresholds in the umbrella state. 
Shaded are the regions where various forms of decay are allowed, see text.
The non-umbrella region for $S\!=\!1/2$ is sketched from Ref.~\cite{yamamoto}, see~\cite{footnote1}. The dot marks 
the values of $\Delta$ and $H$ used in (a) and (b).}
\label{fig_born}
\vskip -0.3cm
\end{figure}

\emph{Model and spectrum.}---%
Owing to frustration and degeneracies of the model, triangular-lattice antiferromagnets in external field 
have a very rich phase diagram \cite{kawamura,korshunov,chubukov91,yamamoto,yamamoto1,eggert}, 
featuring the hallmark plateau, coplanar, and umbrella states, see Ref.~\cite{starykh_review} for a recent review. 
We will focus on the $XXZ$ Hamiltonian with an easy-plane anisotropy whose zero-field ground state is a 120${\degree}$ 
structure
\begin{equation}
\hat{\cal H}=J\sum_{\langle ij\rangle}\left(S^{x}_i S^{x}_j+S^{y}_i S^{y}_j+\Delta S^{z}_i S^{z}_j\right)
-H\sum_{i} S^{z}_i,
\label{model}
\end{equation}
where $\langle ij\rangle$ are nearest-neighbor sites of the triangular lattice, $J\!>\!0$, and $0\!\leq\! \Delta\! <\!1$. 
In an out-of-plane magnetic field, the so-called umbrella structure 
is formed, see Fig.~\ref{fig_linearspectrum}. In the isotropic limit, $\Delta\!=\!1$, the coplanar states are favored instead, 
but $\Delta\!<\!1$ always stabilizes the semiclassical umbrella state  for a range of fields, with the 
$H\!-\!\Delta$ region of its stability for $S\!=\!1/2$ sketched in Fig.~\ref{fig_born}(c) from Ref.~\cite{yamamoto}. 
In Ba$_3$CoSb$_2$O$_9$, estimates of the anisotropy yield $\Delta\!\approx\! 0.9$ \cite{expBaCoSbO2013,expBaCoSbO2016}
with an additional stabilization of the umbrella-like state provided by the interplane coupling  \cite{gekht97,expBaCoSbO2015}.
The linear spin-wave (LSW) treatment of the model (\ref{model}) within the $1/S$-expansion is 
standard, see \cite{supp}. The harmonic magnon energies, $\varepsilon_{\bf k}$, are depicted in 
Fig.~\ref{fig_linearspectrum}  for $S\!=\!1/2$, $\Delta\!=\!0.9$, and $H\!=\!0$ and $H\!=\!0.2 H_s$, 
where $H_s\!=\!6JS(\Delta+1/2)$ is the 
saturation field. The chosen representative field of $0.2H_s$ is within the umbrella region of Fig.~\ref{fig_born}(c) 
and for Ba$_3$CoSb$_2$O$_9$ it corresponds to a modest field of about 6 T \cite{expBaCoSbO2013}. 

In Fig.~\ref{fig_linearspectrum}, one can see the gaps $\propto\!\sqrt{1-\Delta}$  at K and K$^\prime$ points in zero field.
In a finite field, the staggered scalar chirality 
of the umbrella structure, ${\bf S}_i\!\cdot\! ({\bf S}_j\!\times\! {\bf S}_k)$, 
induces inversion symmetry breaking. Because of that, magnon energy acquires 
an asymmetric contribution  \cite{zh96}, $\varepsilon_{\bf k} \neq \varepsilon_{-\bf k}$,
with the energies at K (K$^\prime$) points lowered (raised) proportionally to the field.  
Note that the K and K$'$ points trade their places in 
the domain with a shifted pattern of the 120${\degree}$ order that also corresponds to the flipped
staggered chiralities.
It is 
clear, that the distorted band structure brings down the energy of a minimum of the two-magnon continuum  
associated with the low-energy, roton-like magnons at K-points. 
Given the remaining commensurability of the umbrella state, which retains the 
$3{\bf K}\!=\!0$ property of the 120$\degree$ 
structure, magnon decays may occur in a proximity of the 
K$^\prime$ points via a process 
$\varepsilon_{\bf K'}\Rightarrow\varepsilon_{{\bf K}}+\varepsilon_{{\bf K}(\pm{\bf G}_i)}$, where
${\bf G}_i$'s are the reciprocal lattice vectors. 
While the exact kinematics of such decays is somewhat more complicated, one can simply check where
and at what field the on-shell decay conditions,
$\varepsilon_{{\bf k}}=\varepsilon_{{\bf q}}+\varepsilon_{{\bf k-q}}$, are first met for a given $\Delta$.

This direct verification yields the lower border of the shaded regions in Fig.~\ref{fig_born}(c), which is a union of three 
curves. At large anisotropies, $\Delta\rightarrow 0$, the decay conditions that are fulfilled at the lowest field
are the ones associated with the change of the curvature of the Goldstone mode near the $\Gamma$ point, 
the kinematics familiar from the field-induced decays in the square-lattice \cite{99} and honeycomb-lattice AFs 
\cite{ushoney16}, as well as $^4$He \cite{pitaevskii}. At larger $\Delta$, the threshold field for decays 
is precisely determined by the ``asymmetry-induced'' condition $\varepsilon_{\bf K'}=2\varepsilon_{{\bf K}}$ 
discussed above,
which is given analytically by $H^{*}\!=\!\sqrt{(1-\Delta)/(13-\Delta)}$ and is shown by the dashed line 
in Fig.~\ref{fig_born}(c). Closer to the isotropic
limit, $\Delta\agt 0.7$, the decay conditions are first met away from the high-symmetry points, see some 
discussion of them for the zero-field case and $\Delta>0.92$ in Ref.~\cite{triangle}.

\emph{On(off)-shell decay rate.}---%
To get a sense of the quantitative measure of the 
field-induced broadening effect and of the extent of the affected ${\bf k}$-space, 
we first present the results for the decay rate in the Born approximation
\begin{eqnarray}
\Gamma_{{\bf k}} =
\Gamma_0 \sum_{{\bf q}} \left|\Phi_{{\bf q},{\bf k-q};{\bf k}}\right|^2 
\delta\left(\omega_{{\bf k}}-\omega_{{\bf q}}-\omega_{{\bf k-q}}\right),
\label{eq_gammak}
\end{eqnarray}
where $\Gamma_0\!=\!3\pi J/4$ and $\varepsilon_{{\bf k}}\!=\!3JS\omega_{\bf k}$.
The three-magnon decay vertex $\Phi_{{\bf q},{\bf k-q};{\bf k}}$ is derived from the anharmonic coupling terms of the
$1/S$-expansion of the model (\ref{model}), see \cite{supp}. It combines the effects of noncollinearity 
due to in-plane 120$\degree$ structure and of the field-induced tilting of  spins \cite{99,triangle,RMP}. 
We show  $\Gamma_{{\bf k}}$ for a representative $H\!=\!0.2H_s$ 
and for the same $\Delta\!=\!0.9$ and $S\!=\!1/2$ as above: in Fig.~\ref{fig_born}(a) 
along the MK$^\prime\Gamma$ path (dashed line) and in Fig.~\ref{fig_born}(b) as a 2D intensity plot.

In addition, we also present the results of the self-consistent solution of the off-shell 
Dyson's equation (DE) for $\Gamma_{\bf k}$, in which corrections to the magnon energy are ignored but 
the imaginary part of the the magnon self-energy $\Sigma_{\bf k}(\omega)$ 
due to three-magnon coupling is retained, referred to as the iDE approach:
$\Gamma_{\bf k}\!=\!-\text{Im}\,\Sigma_{\bf k}\left( \varepsilon_{\bf k}+i\Gamma_{\bf k}\right)$. 
This method accounts for a damping of the decaying initial-state magnon and regularizes the 
van Hove singularities associated with the two-magnon continuum that can be seen in the Born results of (\ref{eq_gammak}) 
in Fig.~\ref{fig_born}(a). The same Figure shows the iDE results for $\Gamma_{\bf k}$ (solid line) and the corresponding
magnon spectral function in a lorentzian form (intensity plot). 
We note that the self-consistency schemes that rely
on the broadening of the decay products, such as iSCBA discussed in Refs.~\cite{Mourigal10,Sluckin}, 
are not applicable here because our final-state magnons are well-defined. 
Altogether, our consideration suggests that a significant
$T\!=\!0$ field-induced broadening of quasiparticle peaks due to magnon decays should appear
in a wide vicinity of the K$'$  points in low fields, reaching values of $\Gamma_{\bf k}\!\propto\!0.3J$ (cf. $\alt\!0.1J$ in 
Ba$_3$CoSb$_2$O$_9$ \cite{expBaCoSbO2016}).   

\begin{figure*}
\includegraphics[width=0.99\linewidth]{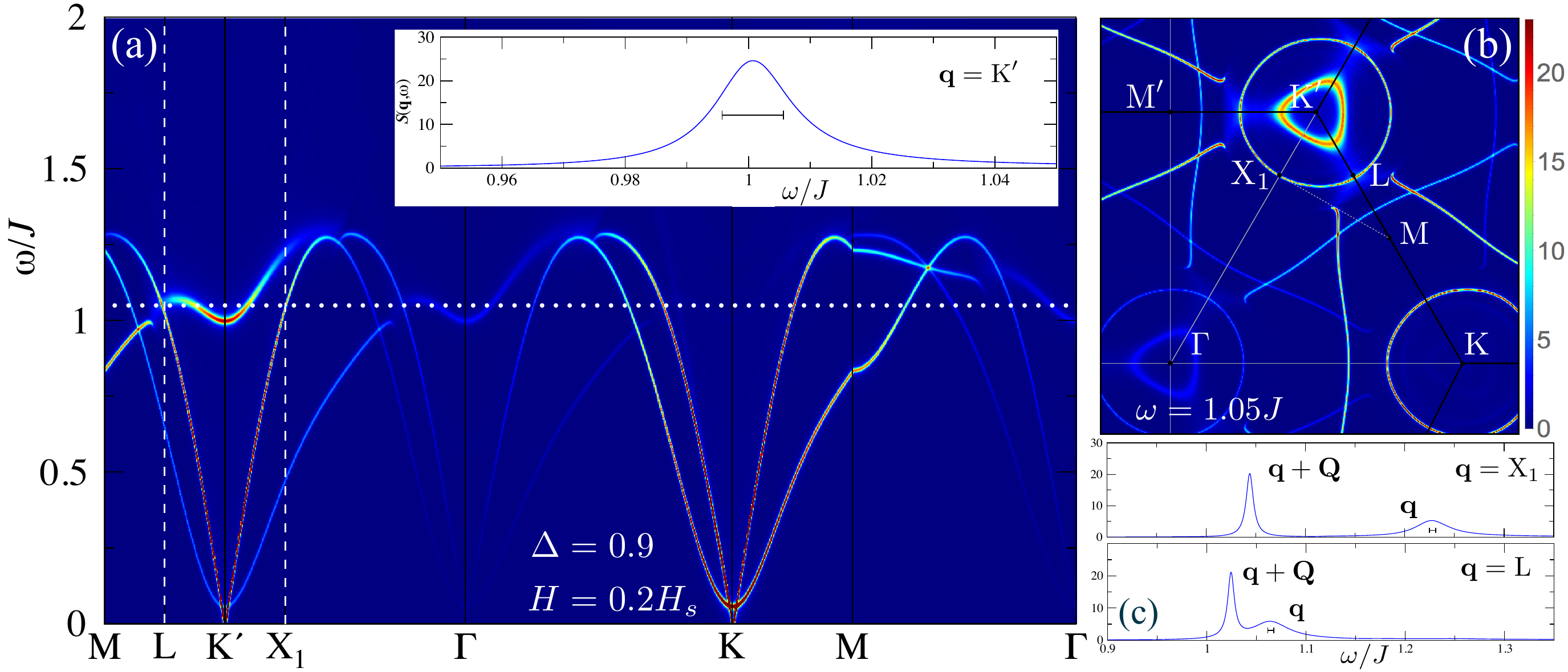}
\caption{(a) Intensity plot of ${\cal S} ({\bf q},\omega)$ along the MK$'\Gamma$KM$\Gamma$ path. 
Dotted and dashed lines are constant-energy cut  in (b) and $\omega$-cuts in (c).
Inset: ${\cal S}({\bf q},\omega)$ vs $\omega$ at K$'$. Bars are artificial width $2\delta\!=\!0.01J$ of the 
calculation \cite{footnote}. 
$S\!=\!1/2$, $\Delta\!=\!0.9$, $H\!=\!0.2H_s$.}
\label{fig_strfac}
\end{figure*}

\emph{Dynamical structure factor.}---%
Next, we evaluate the dynamical spin-spin structure factor ${\cal S} ({\bf q},\omega)$, the quantity directly observed 
in the inelastic neutron scattering experiments. 
Following Ref.~\cite{triSqw}, we approximate ${\cal S} ({\bf q},\omega)$ as a sum of the diagonal terms \cite{supp} of
\begin{equation}
{\cal S}^{\alpha_0 \beta_0} ({\bf q},\omega) =\frac{i}{\pi} \text{Im} \int_{-\infty}^{\infty} dt e^{i\omega t} \langle T S^{\alpha_0}_{{\bf q}}(t) S^{\beta_0}_{-{\bf q}}(0) \rangle.
\label{eq_strfac}
\end{equation}
Transforming to the local (rotating) reference frame of the ordered moments and keeping terms that contribute 
to the leading $1/S$ order \cite{triSqw} yields 
\begin{eqnarray}
{\cal S}^{x_0 x_0}({\bf q},\omega)=\frac{1}{4}\Big[{\cal S}^{yy}_{{\bf q}+}+{\cal S}^{y y}_{{\bf q}-}
+2i\sin\theta\left({\cal S}^{x y}_{{\bf q}+}-{\cal S}^{x y}_{{\bf q}-}\right)\nonumber\\
+\sin^2 \theta \left({\cal S}^{x x}_{{\bf q}+}+{\cal S}^{x x}_{{\bf q}-}\right) +\cos^2 \theta \left({\cal S}^{zz}_{{\bf q}+}+{\cal S}^{z z}_{{\bf q}-}\right)\Big],
\label{eq_Swq1}
\\
{\cal S}^{z_0 z_0}({\bf q},\omega)=\cos^2 \theta \ {\cal S}^{x x}_{{\bf q}}+\sin^2\theta \  {\cal S}^{z z}_{{\bf q}}, 
 \ \quad\quad\quad\quad\quad\quad\nonumber
\end{eqnarray}
and ${\cal S}^{y_0 y_0}({\bf q},\omega)\!=\!{\cal S}^{x_0 x_0}({\bf q},\omega)$. 
Here we used the antisymmetric nature of the $xy$ contribution 
${\cal S}^{x y}_{{\bf q}}\!=\!-{\cal S}^{yx}_{{\bf q}}$ and introduced 
shorthand notations for ${\cal S}^{\alpha\beta}_{\bf q}\!\equiv\! {\cal S}^{\alpha\beta}({\bf q},\omega)$ 
and ``shifted'' momenta ${\bf q}\pm\equiv {\bf q}\pm{\bf K}$, with $\theta$ being the out-of-plane canting angle 
of  spins. In the local reference frame,  ${\cal S}^{zz}_{\bf q}$ components of the dynamical structure factor 
are ``longitudinal'', i.e., are due to the two-magnon continuum, having no sharp  quasiparticle features \cite{triSqw}.
The rest of Eq.~(\ref{eq_Swq1}) is ``transverse'', i.e., is related to the single-magnon spectral function,
${\cal S}^{x(y)x(y)}_{\bf q}\!\propto\!A({\bf q},\omega)$,
with different kinematic ${\bf q}$-dependent formfactors, where
$A({\bf q},\omega)\!=\!-(1/\pi){\rm Im} G({\bf q},\omega)$ and the diagonal magnon 
Green's function is $G({\bf q},\omega)=[\omega-\varepsilon_{\bf q}-\Sigma_{\bf q}(\omega)+i\delta]^{-1}$. 
Thus, the dynamical structure factor of  the $XXZ$ TLAF in a field should feature three
overlapping single-magnon spectral functions,  $A({\bf q},\omega)$ and $A({\bf q}\pm {\bf K},\omega)$, 
with different weights according to (\ref{eq_Swq1}) and \cite{supp}, see also \cite{footnote2}.

In our consideration, we  include all 
contributions to the one-loop magnon self-energy $\Sigma_{\bf q}(\omega)$ of the 
$1/S$-order of the non-linear spin-wave theory \cite{RMP}.
Namely, there are two more terms in addition to decay diagram: 
the source diagram and the Hartree-Fock correction, the latter comprised of the contributions from the 
four-magnon interactions (quartic terms) and from the quantum corrections to the 
out-of-plane canting angle  of spins, see \cite{supp} for technical details,
\begin{equation}
\Sigma_{\bf q}(\omega)=\Sigma^{\rm HF}_{\bf q}+\Sigma_{\bf q}^{d} (\omega)+\Sigma_{\bf q}^{s} (\omega).
\end{equation}
Having included all one-loop contributions also allows us to  consistently take into account the $\omega$-dependence 
of the magnon spectral function. Below we demonstrate that  
anharmonic interactions lead to broadening of magnon quasiparticle peaks, redistribution of spectral weight, and other
dramatic changes in the spectrum. 
 
In Fig.~\ref{fig_strfac}, we present our results for the dynamical structure factor ${\cal S} ({\bf q},\omega)$ 
in (\ref{eq_Swq1}) of the model (\ref{model}) for $S\!=\!1/2$, $\Delta\!=\!0.9$, and $H\!=\!0.2H_s$. 
First, there is a strong downward bandwidth renormalization by about  
30\% compared to the LSW results in Fig.~\ref{fig_linearspectrum}, which is characteristic to the TLAFs 
\cite{zheng06,starykh06,triangle}.
The most important result is a significant broadening of magnon spectra for an extensive range of momenta, 
accompanied by well-pronounced termination points with distinctive bending of spectral lines 
\cite{Plumb} and other
non-Lorentzian features that are associated with crossings of the two-magnon continuum.
The broadening can be seen in a wide proximity of the K$'$ points of the Brillouin zone as 
well as in the equivalent regions of the ``$\pm {\bf K}$-shifted'' components of the structure factor. Despite the 
strong renormalization of the spectrum, the extent of the affected ${\bf q}$-region is about the same as in the
on-shell consideration in Fig.~\ref{fig_born}. 

The inset of Fig.~\ref{fig_strfac}(a) shows ${\cal S} ({\bf q},\omega)$ vs $\omega$
at a representative K$'$ point that exhibits a modest broadening 
compared with the artificial width ($2\delta$) of the calculation. The $\omega$-cuts at the 
X$_1$ and L points near the boundaries of the decay region in Fig.~\ref{fig_strfac}(c) 
show much heavier damping in one 
of the component of ${\cal S} ({\bf q},\omega)$, which coexists with the well-defined spectral peak  
from the ``shifted'' component. The enhancement of magnon decays near the edge of  decay region 
also correlates well with the on-shell results  in Fig.~\ref{fig_born} and points to the van 
Hove singularities of the two-magnon continuum as a culprit. 
The 2D intensity map of the constant-energy cut of ${\cal S} ({\bf q},\omega)$ at $\omega\!=\!1.05J$ 
is shown in Fig.~\ref{fig_strfac}(b), where one can see multiple signatures of 
the broadening, spectral weight redistribution around K$'$, and termination points. 

\begin{figure}
\includegraphics[width=0.99\linewidth]{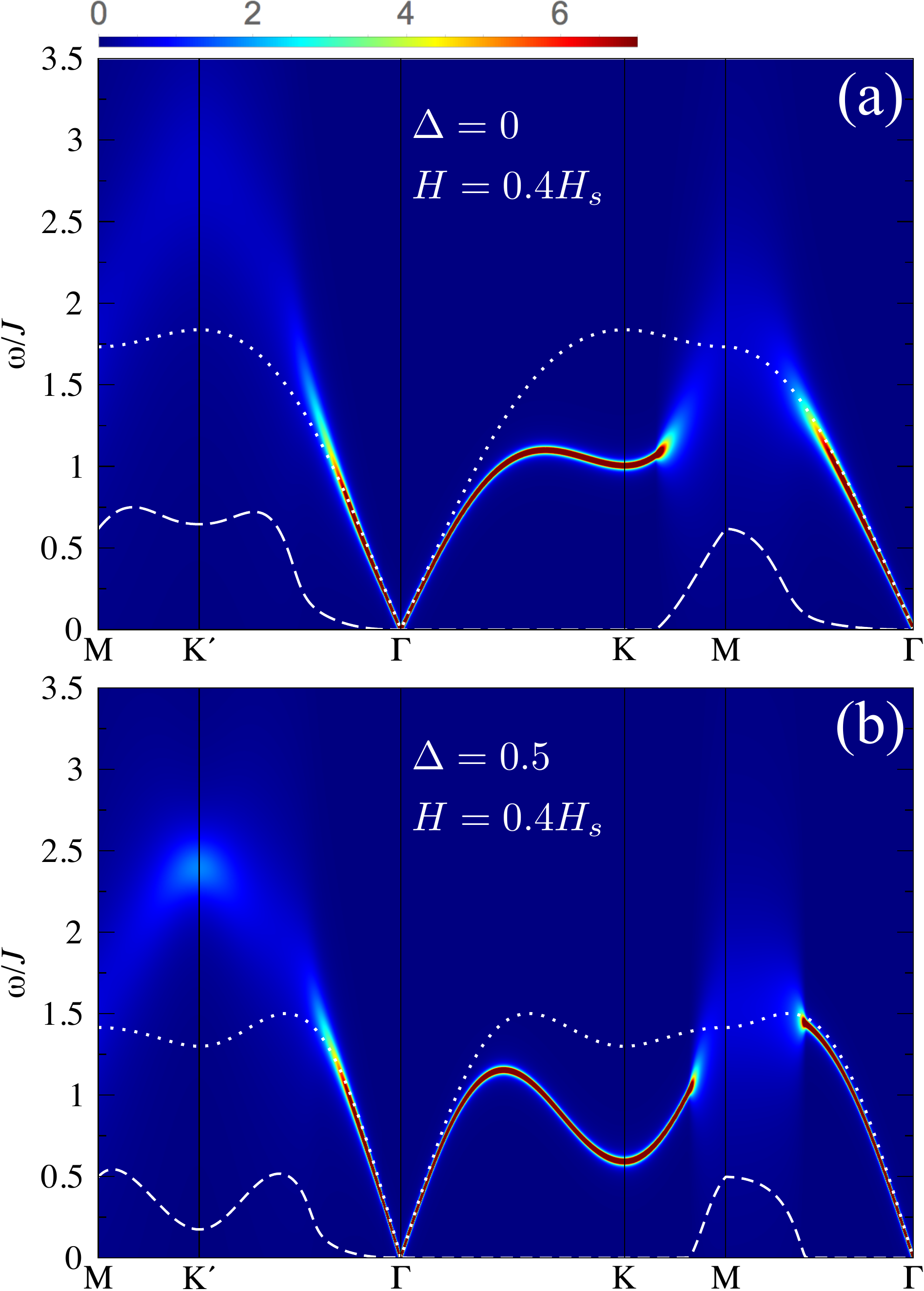}
\caption{Intensity plots of the spectral function with the iDE $\Gamma_{\bf k}$ (dashed lines)
for $S\!=\!1/2$, $H\!=\!0.4H_s$, and $\Delta\!=\!0$ in (a) and $\Delta\!=\!0.5$ in (b). 
Dotted lines are the LSW spectra for $H\!=\!0$.}
\label{fig_sfd0}
\end{figure}

\emph{Larger anisotropy.}---%
We complement our consideration of the model (\ref{model}) by demonstrating the effects 
of magnetic field on the magnon spectrum for the TLAFs with large easy-plane anisotropy. 
In the strongly-anisotropic limit, $\Delta\!=\!0$, the non-linear anharmonic coupling of magnons is known 
to result in a very strong spectrum renormalization (about 50\%), but with no decays  kinematically 
allowed \cite{triangle}. For $S\!=\!1/2$ and small enough $\Delta$, Born approximation 
and the $1/S$, one-loop, $\omega$-dependent self-energy approach are somewhat inconsistent in that 
the first produces unphysically large $\Gamma_{\bf k}$ for $H\!\agt\! 0.3H_s$ and the second shows strong 
spectrum renormalization that avoids decays for $H\!\alt\! 0.5H_s$. 
Since the reason for this discrepancy is  the lack of self-consistency, we resort to 
the (partially) self-consistent iDE approach described above. In Fig.~\ref{fig_sfd0}, we show its 
results for the magnon spectral function with the Lorentzian broadening $\Gamma_{\bf k}$ 
for $\Delta\!=\!0$ and $\Delta\!=\!0.5$ and for $H\!=\!0.4H_s$. What is remarkable is not 
only a persistent pattern of a wide ${\bf k}$-region of the strongly overdamped  high-energy magnons 
[cf., Fig.~\ref{fig_born}(a)], but also the magnitudes of their broadening, which reach the values of almost a
half of the magnon bandwidth even after a self-consistent regularization. 

\emph{Conclusions.}---%
We have provided a detailed analysis of the field-induced dynamical response of the $XXZ$ model on the 
triangular lattice within the umbrella phase. We have demonstrated a ubiquitous presence of 
significant damping of the high-energy magnons already in moderate fields, $H\agt 0.2H_s$.
Other characteristic features, such as significant spectral weight redistribution and termination points that separate 
well-defined excitations from the ones that are overdamped, are also expected to occur.
The key physical ingredients of this dramatic spectral transformation are 
a strong spin noncollinearity, which is retained by the umbrella state 
and is essential for the anharmonic magnon coupling and decays, and the tilted, ${\bf k}\leftrightarrow -{\bf k}$ 
asymmetric magnon band structure, owing its origin to the staggered chirality of the umbrella state that 
breaks the inversion symmetry. Our consideration pertains in particular to Ba$_3$CoSb$_2$O$_9$,
which is currently a prime candidate for observing aforementioned properties in reasonably small fields
reachable in experimental setup.
Our work should be of a qualitative  and quantitative guidance for observations of the
dynamical structure factor in the inelastic neutron-scattering experiments
in this and other related systems.

\begin{acknowledgments}
\emph{Acknowledgments.}---%
We acknowledge useful conversations with Martin Mourigal and Cristian Batista. We are particularly indebted 
to Martin for his unbiased experimental intuition that led to \cite{footnote2}.
This work was supported by the U.S. Department of Energy,
Office of Science, Basic Energy Sciences under Award \# DE-FG02-04ER46174.
A. L. C. would like to thank the Kavli Institute for Theoretical Physics where part of this work was done. 
The work at KITP was supported in part by NSF Grant No. NSF PHY11-25915.
\end{acknowledgments}

\vskip -0.3cm


\newpage
\onecolumngrid
\begin{center}
{\large\bf Field-induced decays in $XXZ$ triangular-lattice antiferromagnets: \\ Supplemental Material}\\ 
\vskip0.35cm
P. A. Maksimov$^1$, M. E. Zhitomirsky$^2$, and A. L. Chernyshev$^1$  \\
\vskip0.15cm
{\it \small $^1$Department of Physics and Astronomy, University of California, Irvine, California
92697, USA}\\
{\it \small $^2$CEA, INAC-PHELIQS, F-38000, Grenoble, France}\\
{\small (Dated: September 19, 2016)}\\
\vskip 0.1cm \
\end{center}
\twocolumngrid

\setcounter{equation}{0}
\setcounter{figure}{0}

Here we present the details of the nonlinear spin-wave formalism for the $XXZ$ model 
on a triangular lattice in an out-of-plane magnetic field and in the semi-classical umbrella state. 
The formalism bears significant similarities to the Heisenberg triangular antiferromagnet case in zero field 
\cite{s_starykh06,s_triangle} and to the square-lattice antiferromagnet in a field \cite{s_99}, with several details that 
differ from both.

\subsection{Model and spin transformation}
\begin{figure}[b]
\includegraphics[width=\linewidth]{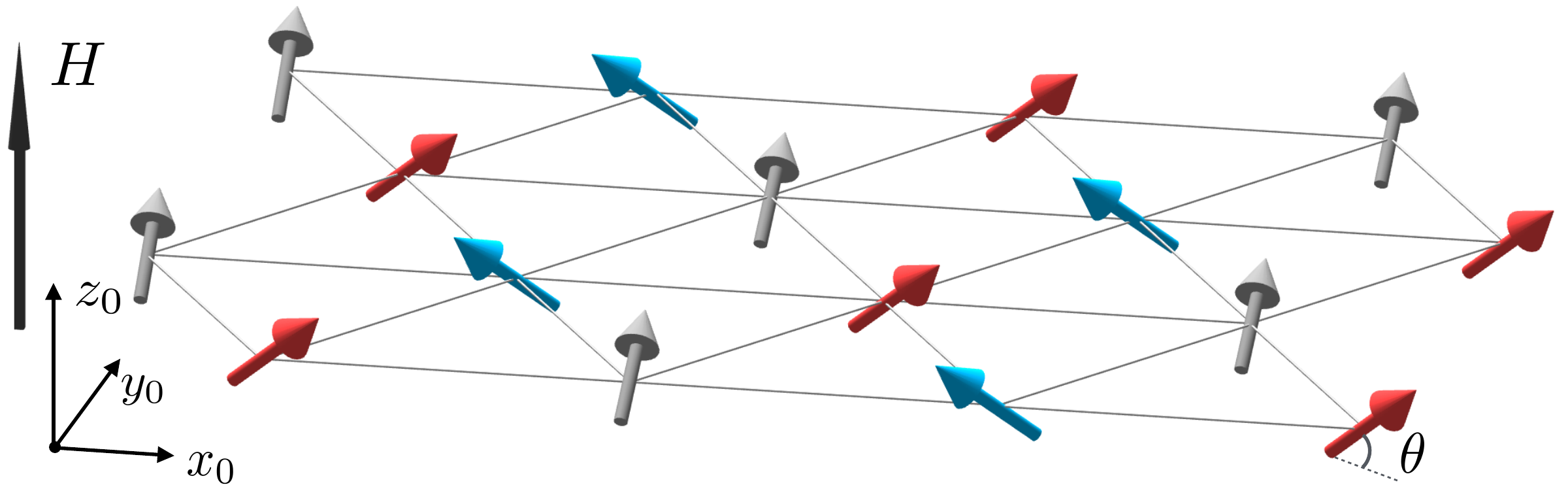}
\caption{Umbrella spin structure on the triangular lattice, $\theta$ is the out-of-plane canting angle.}
\label{s_fig_umbrella}
\end{figure}

The nearest-neighbor $XXZ$ Hamiltonian on a triangular lattice in external out-of-plane field is
\begin{eqnarray}
\hat{\cal H}=J\sum_{\langle ij\rangle}\Big({\bf S}_i \cdot {\bf S}_j-\left(1-\Delta\right) S^{z_0}_i S^{z_0}_j\Big)
-H\sum_{i} S^{z_0}_i\ ,
\label{s_eq_hamiltonian1}
\end{eqnarray}
where  the sum is over the nearest-neighbor bonds $\langle ij\rangle$, $J\!>\!0$ is  an exchange coupling constant, 
$0\!\leq\Delta\!\leq\!1$ is the easy-plane anisotropy parameter, and $H$ is external magnetic field in units of $g \mu_B$.  
The ground state in zero field is a 120$^\circ$ structure with the ordering vector $\textbf{Q}\!=\!\left(\frac{4\pi}{3},0\right)$.
In applied field, spins cant towards the field direction to form the umbrella structure shown in 
Fig.~\ref{s_fig_umbrella}. Thus, we need to align the local spin-quantization axis 
on each site in the direction given by the spin configuration,
with the canting angle defined from the energy minimization. 

The corresponding general transformation of the spin components from the laboratory reference frame 
$\{x_0,y_0,z_0\}$ to the local reference frame $\{x,y,z\}$ can be performed using two consequent rotations
\begin{eqnarray}
{\bf S}^0_i={\bf R}_{{\bf Q}}\cdot {\bf R}_{\theta} \cdot{\bf S}_i\, ,
\label{s_eq_frametransform}
\end{eqnarray}
where the matrix ${\bf R}_{{\bf Q}}$ does  rotations in the  $\{x_0y_0\}$-plane
\begin{eqnarray}
{\bf R}_{{\bf Q}}=
\left( \begin{array}{ccc} 
\cos \varphi_i & -\sin\varphi_i & 0\\ 
\sin\varphi_i & \cos\varphi_i & 0\\
0 & 0 & 1 
\end{array}\right),
\label{s_Ri}
\end{eqnarray}
where $\varphi_i\!=\!{\bf Q} \cdot {\bf r}_i$ is an in-plane angle. The out-of-plane
rotation is done by ${\bf R}_{\theta}$ within the $\{x_0,z_0\}$-plane 
\begin{eqnarray}
{\bf R}_{\theta}=\left( \begin{array}{ccc} \sin \theta& 0 & \cos \theta \\ 
0 & 1 & 0\\
-\cos\theta & 0 & \sin\theta \end{array}\right),
\label{s_Rth}
\end{eqnarray}
where $\theta$ is the out-of-plane canting angle. The full transformation is explicitly given by
\begin{align}
S^{x_0} &= S^x \sin\theta \cos {\bf Q} \cdot {\bf r} - S^{y} \sin {\bf Q} \cdot {\bf r}
+ S^{z} \cos\theta \cos {\bf Q} \cdot {\bf r}, \nonumber\\
S^{y_0} &= S^x \sin\theta \sin {\bf Q} \cdot {\bf r} 
+ S^{y} \cos {\bf Q} \cdot {\bf r}+ S^{z} \cos\theta \sin {\bf Q} \cdot {\bf r}, \nonumber\\
S^{z_0} &= -S^x \cos\theta +S^{z} \sin \theta.
\end{align}

\subsection{Transformed Hamiltonian}

After performing the axis-rotation transformation (\ref{s_eq_frametransform}), we split Hamiltonian 
into ``even'' and ``odd'' parts
\begin{eqnarray}
\hat{\cal H}=\hat{\cal H}_{\rm even}+\hat{\cal H}_{\rm odd}\, ,
\end{eqnarray}
which   become even and odd in bosonic field operators
\begin{eqnarray}
&&\hat{\cal H}_{\rm even}=J\sum_{\langle ij \rangle} S^{x}_i S^{x}_j \left[ \sin^2 \theta \cos \delta \varphi_{ij} +\Delta \cos^2 \theta\right]\nonumber\\
&&\quad\quad\quad\quad\quad\quad+S^{y}_i S^{y}_j \cos \delta \varphi_{ij}\label{s_eq_Heven}\\
&&\quad\quad\quad\quad\quad\quad
+S^{z}_i S^{z}_j \left[ \cos^2 \theta \cos \delta \varphi_{ij}+\Delta \sin^2 \theta\right]\nonumber\\
&&\quad\quad\quad\quad\quad\quad
+ \left( S^{x}_i S^{y}_j - S^{y}_i S^{x}_j \right) \sin \delta \varphi_{ij} \sin \theta \nonumber\\
&&\quad\quad\quad\quad\quad\quad- H\sum_i S^{z}_i \sin \theta .\nonumber
\end{eqnarray}
Within the $1/S$ expansion, this term will yield classical energy, harmonic spectrum, and four-magnon interactions. 
The three-magnon interactions will be produced by the odd Hamiltonian and originate from the non-collinear spin 
structure 
\begin{eqnarray}
&&\hat{\cal H}_{\rm odd}=J \sum_{\langle ij \rangle}  \left( S^{x}_i S^{z}_j + S^{z}_i S^{x}_j\right) \sin \theta \cos \theta \left(-\Delta+\cos\delta \varphi_{ij} \right) \nonumber\\
&&\quad\quad\quad\quad\quad\quad
+ \left( S^{z}_i S^{y}_j - S^{y}_i S^{z}_j \right) \cos\theta \sin \delta \varphi_{ij} \nonumber\\
&&\quad\quad\quad\quad\quad\quad+ \sum_i H S^{x}_i \cos \theta,
\label{s_eq_Hodd}
\end{eqnarray}
where $\delta \varphi_{ij}= \varphi_i-\varphi_j= \pm 120^{\degree}$, i.e., the 
in-plane spin configuration is retained in the umbrella state.
The classical energy can be obtained from (\ref{s_eq_Heven}) as
\begin{eqnarray}
\frac{E_{\rm cl}}{NJS^2}=
3\Big(\Delta\sin^2 \theta-\frac{1}{2}\cos^2\theta\Big)
-\frac{H\sin\theta}{JS}\, ,
\label{s_Ecl}
\end{eqnarray}
with the minimization yielding a relation between the field $H$ and  the canting angle $\theta$ as:  
$H\!=\!H_s \sin \theta$ where $H_s\!=\!6JS\left(\Delta+ \frac{1}{2} \right)$ is the saturation field. 
The subsequent treatment of the spin Hamiltonian in (\ref{s_eq_Heven}) and (\ref{s_eq_Hodd}) 
involves a standard Holstein-Primakoff transformation 
\begin{equation}
S^{+}_i=a^{\phantom{\dagger}}_i\sqrt{2S-a^\dagger_i a^{\phantom{\dagger}}_i}, \quad S^z_i=S-a^\dagger_i a^{\phantom{\dagger}}_i.
\label{s_HP}
\end{equation}

\subsection{Linear spin-wave theory}

\begin{figure}
\includegraphics[width=\linewidth]{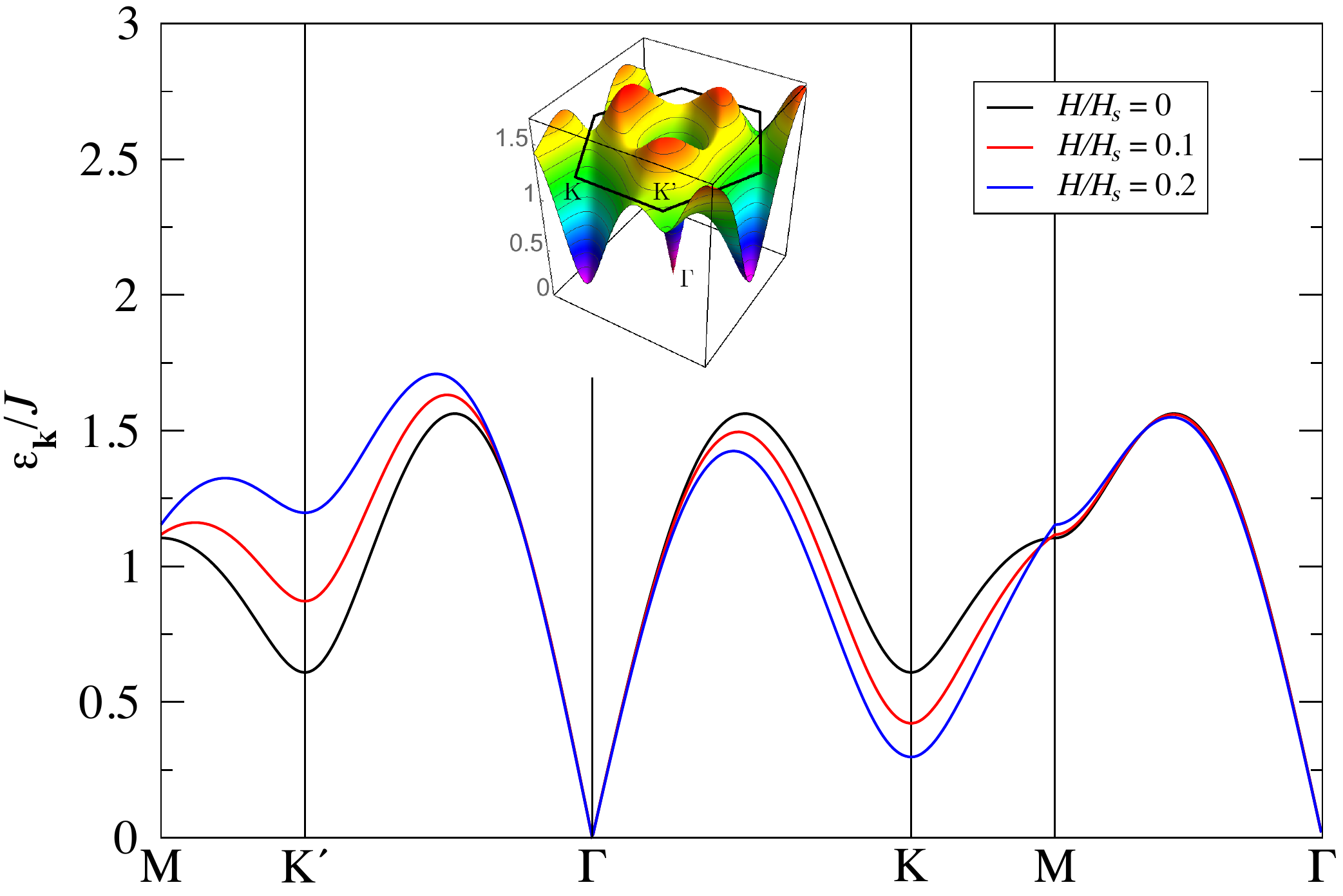}
\caption{Magnon linear spin-wave spectrum for $\Delta=0.9$ and several fields. 
Inset: 3D plot of $\varepsilon_{\bf k}$ for $H=0.2 H_s$.}
\label{s_fig_spectrum}
\end{figure}

The next non-vanishing term in the $1/S$ expansion of (\ref{s_eq_Heven}) 
beyond $E_{\rm cl}$ is the quadratic Hamiltonian
\begin{eqnarray}
&&\hat{\cal H}^{(2)}=JS  \sum_{\langle ij\rangle}\bigg[a^\dagger_{i} a^{\phantom{\dag}}_{i} 
+\left(\frac{2\lambda-1}{4}\right)\left(a^\dagger_{i} a^{\phantom{\dag}}_{j}+a^\dagger_{i} a^{\phantom{\dag}}_{j}\right)
\nonumber\\
&&\phantom{\hat{\cal H}^{(2)}=JS  \sum_{\langle ij\rangle}}
-i \left(a^\dagger_{i} a^{\phantom{\dag}}_{j}-a^\dagger_{j} a^{\phantom{\dag}}_{i}\right)
   \sin \delta \varphi_{ij}  \sin \theta\nonumber\\
&&\phantom{\hat{\cal H}^{(2)}=JS  \sum_{\langle ij\rangle}}
+\left(\frac{2\lambda+1}{4}\right)\left(a^{\phantom{\dag}}_{i} a^{\phantom{\dag}}_{j} 
+a^\dagger_{i} a^\dagger_{j}\right)\bigg],
\label{s_eq_H2}
\end{eqnarray}
where $\lambda=\left( \Delta+\frac{1}{2}\right) \cos^2 \theta-\frac{1}{2}$. Next, we introduce  Fourier transformation
\begin{eqnarray}
a^{\phantom{\dag}}_{i}=
\frac{1}{\sqrt{N}} \sum_{{\bf k}} e^{i{\bf k} {\bf r}_i} 
a^{\phantom{\dag}}_{\bf k}\, .
\label{s_eq_fourier}
\end{eqnarray}
Reciprocal vectors of triangular lattice are
\begin{equation}
\mathbf{b}_1=\left( 0,\frac{4\pi}{\sqrt{3}} \right),~\mathbf{b}_2=\left( 2\pi,-\frac{2\pi}{\sqrt{3}} \right).
\end{equation}
This gives the harmonic Hamiltonian
\begin{equation}
\hat{\cal H}^{(2)}=3JS\sum_{{\bf k}} 
\left(A_{\bf k} +C_{\bf k}\right) a^\dagger_{{\bf k}}a^{\phantom{\dag}}_{{\bf k}}   
-\frac{B_{\bf k}}{2} \left( a^\dagger_{{\bf k}} a^{\dagger}_{-{\bf k}}
+{\rm H.c.}\right), \label{s_eq_H2k}
\end{equation}
with the parameters
\begin{eqnarray}
&&A_{\bf k}=1+\gamma_{{\bf k}}\left[ \left( \Delta+\frac{1}{2}\right)\cos^2 \theta-1 \right] , \\
&&B_{\bf k}=-\gamma_{{\bf k}}\left( \Delta+\frac{1}{2}\right)\cos^2 \theta ,\\
&&C_{\bf k}=\sqrt{3} \bar{\gamma}_{{\bf k}} \sin\theta\, .
\label{s_eq_ck}
\end{eqnarray}
Here $\gamma_{{\bf k}}$ and $\bar{\gamma}_{{\bf k}}$ are the nearest-neighbor amplitudes 
\begin{eqnarray}
\gamma_{{\bf k}}=\frac{1}{6}\sum_{{\bm \delta}_i} e^{i{{\bf k}}\cdot {\bm \delta}_i},
\  \bar{\gamma}_{{\bf k}}=\frac{1}{6}\sum_{{\bm \delta}_i} \mbox{sign}(\sin\delta\varphi_{ij})\,  
e^{i{{\bf k}}\cdot {\bm \delta}_i},
\label{s_gk}
\end{eqnarray}
or, explicitly,
\begin{eqnarray}
&&\gamma_{{\bf k}}=\frac{1}{3}\Big(\cos k_x+2\cos\frac{k_x}{2}\cos \frac{\sqrt{3}k_y}{2}\Big),\\
&&\bar{\gamma}_{{\bf k}}=\frac{1}{3} \Big(\sin k_x - 2\sin \frac{k_x}{2} \cos \frac{\sqrt{3}k_y}{2}\Big).
\end{eqnarray}
The Bogolyubov transformation of (\ref{s_eq_H2})
\begin{eqnarray}
c_{{\bf k}}=u_{{\bf k}} a^{\phantom{\dag}}_{{\bf k}}+v_{{\bf k}} a^{\dagger}_{-{\bf k}}\, ,
\label{s_eq_bogolyubov}
\end{eqnarray}
is standard with the parameters given by
\begin{eqnarray}
2u_{{\bf k}}v_{{\bf k}}=\frac{B_{{\bf k}}}{\sqrt{A^2_{{\bf k}}-B^2_{{\bf k}}}}\, ,\quad
u^2_{{\bf k}}+v^2_{{\bf k}}=\frac{A_{{\bf k}}}{\sqrt{A^2_{{\bf k}}-B^2_{{\bf k}}}}\, .
\label{s_eq_bogolyubovcoefficients}
\end{eqnarray}
 Finally, the excitation spectrum is 
\begin{equation}
\varepsilon_{{\bf k}}\!=\!3JS \omega_{{\bf k}},
\label{s_eq_spectrum}
\end{equation}
where
$\omega_{{\bf k}}\!=\!\sqrt{A^2_{{\bf k}}-B^2_{{\bf k}}}+C_{\bf k}$. Note that because $C_{-\bf k}\!=\!-C_{\bf k}$, it 
is not affected by Bogolyubov transformation and $u_{\bf k}$ and $v_{\bf k}$ remain even
under ${\bf k}\!\rightarrow\! -{\bf k}$. Fig.~\ref{s_fig_spectrum} shows the linear spin-wave theory spectra 
in the range of parameters applicable to Ba$_3$CoSb$_2$O$_9$. Since the field-induced 
staggered scalar chirality of the umbrella structure, ${\bf S}_i\cdot ({\bf S}_j\times {\bf S}_k)$,  
breaks the inversion symmetry, it leads to an asymmetry of the spectrum \cite{s_zh96}, 
$\varepsilon_{\bf k}\! \neq \!\varepsilon_{-\bf k}$.
Most importantly, it shifts magnon energy at K and K$'$ corners of the Brillouin zone in the opposite directions.

\subsection{Cubic vertices}

Holstein-Primakoff transformation in \eqref{s_eq_Hodd} yields the three-magnon interaction 
\begin{eqnarray}
&& \hat{\cal H}^{(3)}=J\sqrt{\frac{S}{2}}\sum_{\langle ij\rangle} \sin 2\theta \left( \Delta+\frac{1}{2}\right) (a^\dagger_i + a^{\phantom{\dag}}_i) a^\dagger_j a^{\phantom{\dag}}_j\nonumber\\
&& \phantom{\hat{\cal H}^{(3)}=J\sqrt{\frac{S}{2}}\sum_{\langle ij\rangle}}
-2i\cos\theta  \sin \delta\varphi_{ij} \, a^\dagger_i a^{\phantom{\dag}}_i(a^\dagger_j - a^{\phantom{\dag}}_j) , \ \ \  \ \ \ \ 
\label{s_eq_H3ij}
\end{eqnarray}
where the first term is due to the out-of-plane spin noncollinearity and  the second is due to the 120$^\circ$ in-plane order.
The Fourier transformation of (\ref{s_eq_H3ij}) yields
\begin{eqnarray}
&&\hat{\cal H}^{(3)}= 3J \sqrt{\frac{S}{2}}\sum_{{\bf k},{\bf q}}
\left[ \gamma_{{\bf k}}\left( \Delta +\frac{1}{2}\right)\sin 2\theta - \sqrt{3} \bar{\gamma}_{{\bf k}} \cos \theta \right]
\nonumber\\ 
&&\phantom{\hat{\cal H}^{(3)}= 3J \sqrt{\frac{S}{2}}\sum_{{\bf k},{\bf q}}}
\times  \left(
 a^{\dagger}_{{\bf q}} a^{\dagger}_{{\bf k}} 
a^{\phantom{\dag}}_{{\bf k}+{\bf q}}+ {\rm H.c.}\right). 
\label{s_eq_H3k}
\end{eqnarray}
Finally, the Bogolyubov transformation (\ref{s_eq_bogolyubov}) of (\ref{s_eq_H3k}) yields the cubic Hamiltonian for the 
magnon eigenmodes
\begin{eqnarray}
&&\hat{\cal H}^{(3)}=\frac{3J}{3!} \sqrt{\frac{S}{2}}\sum_{-{\bf p}={\bf k}+{\bf q}} \left(
\Xi_{{\bf q}{\bf k}{\bf p}} 
c^{\dagger}_{{\bf q}} c^{\dagger}_{{\bf k}} c^{\dagger}_{{\bf p}}+{\rm H.c.}\right),
\label{s_Hsource}
\\
&&\phantom{\hat{\cal H}^{(3)}}
+\frac{3J}{2!} \sqrt{\frac{S}{2}}\sum_{-{\bf p}={\bf k}+{\bf q}} \left(
\Phi_{{\bf q}{\bf k};{\bf p}} 
c^{\dagger}_{{\bf q}} c^{\dagger}_{{\bf k}} c^{\phantom{\dag}}_{-{\bf p}}+{\rm H.c.}\right),\quad\quad
\label{s_Hdecay}
\end{eqnarray}
where the combinatorial factors are due to symmetrization in the source (\ref{s_Hsource}) and decay (\ref{s_Hdecay}) vertices
given by
\begin{eqnarray}
&&{\Xi}_{{\bf q}{\bf k}{\bf p}} = 
F_{{\bf q}} (u_{{\bf q}}+v_{{\bf q}})
(u_{{\bf k}}v_{{\bf p}}+v_{{\bf k}}u_{{\bf p}})\nonumber\\
&&\phantom{{\Xi}_{{\bf q}{\bf k}{\bf p}}}
+ F_{{\bf k}}
(u_{{\bf k}}+v_{{\bf k}})
(u_{{\bf q}}v_{{\bf p}}+v_{{\bf q}}u_{{\bf p}})\nonumber\\
&&\phantom{{\Xi}_{{\bf q}{\bf k}{\bf p}}}
+ F_{{\bf p}}
(u_{{\bf p}}+v_{{\bf p}})
(u_{{\bf q}}v_{{\bf k}}+v_{{\bf q}}u_{{\bf k}})\label{s_eq_sourcevertex}\\
&&\phantom{{\Xi}_{{\bf q}{\bf k}{\bf p}}}
+\bar{F}_{{\bf q}} (u_{{\bf q}}-v_{{\bf q}})
(u_{{\bf k}}v_{{\bf p}}+v_{{\bf k}}u_{{\bf p}})\nonumber\\
&&\phantom{{\Xi}_{{\bf q}{\bf k}{\bf p}}}
+ \bar{F}_{{\bf k}}
(u_{{\bf k}}-v_{{\bf k}})
(u_{{\bf q}}v_{{\bf p}}+v_{{\bf q}}u_{{\bf p}})\nonumber\\
&&\phantom{{\Xi}_{{\bf q}{\bf k}{\bf p}}}
+ \bar{F}_{{\bf p}}
(u_{{\bf p}}-v_{{\bf p}})
(u_{{\bf q}}v_{{\bf k}}+v_{{\bf q}}u_{{\bf k}})\, ,\nonumber\\
&&{\Phi}^{\eta\nu\mu}_{{\bf q}{\bf k};{\bf p}} =  
F_{{\bf q}}
(u_{{\bf q}}+v_{{\bf q}})
(u_{{\bf k}}u_{{\bf p}}+v_{{\bf k}}v_{{\bf p}})\nonumber\\
&&\phantom{{\Phi}_{{\bf q}{\bf k};{\bf p}}}
+ F_{{\bf k}}
(u_{{\bf k}}+v_{{\bf k}})
(u_{{\bf q}}u_{{\bf p}}+v_{{\bf q}}v_{{\bf p}})\nonumber\\
&&\phantom{{\Phi}_{{\bf q}{\bf k};{\bf p}}}
+ F_{{\bf p}}
(u_{{\bf p}}+v_{{\bf p}})
(u_{{\bf q}}v_{{\bf k}}+v_{{\bf q}}u_{{\bf k}}) \label{s_eq_decayvertex}
\\
&&\phantom{{\Phi}_{{\bf q}{\bf k};{\bf p}}}
+\bar{F}_{{\bf q}}
(u_{{\bf q}}-v_{{\bf q}})
(u_{{\bf k}}u_{{\bf p}}+v_{{\bf k}}v_{{\bf p}})
\nonumber\\
&&\phantom{{\Phi}_{{\bf q}{\bf k};{\bf p}}}
+ \bar{F}_{{\bf k}}
(u_{{\bf k}}-v_{{\bf k}})
(u_{{\bf q}}u_{{\bf p}}+v_{{\bf q}}v_{{\bf p}})\nonumber\\
&&\phantom{{\Phi}_{{\bf q}{\bf k};{\bf p}}}
- \bar{F}_{{\bf p}}
(u_{{\bf p}}-v_{{\bf p}})
(u_{{\bf q}}v_{{\bf k}}+v_{{\bf q}}u_{{\bf k}}), \nonumber\quad\quad
\end{eqnarray}
with  ${\bf k}\leftrightarrow -{\bf k}$ symmetric and antisymmetric amplitudes 
\begin{eqnarray}
F_{\bf k}&=&\gamma_{{\bf k}}\left( \Delta +\frac{1}{2}\right)\sin 2\theta, \\
\bar{F}_{\bf k}&=&-\sqrt{3} \bar{\gamma}_{{\bf k}} \cos \theta.
\end{eqnarray}
The terms with $\bar{F}_{\bf k}$ in the vertices have a structure familiar from the zero-field consideration
\cite{s_triangle}, while the symmetric terms are from the out-of-plane canting. 

\begin{figure}[t]
\includegraphics[width=0.99\linewidth]{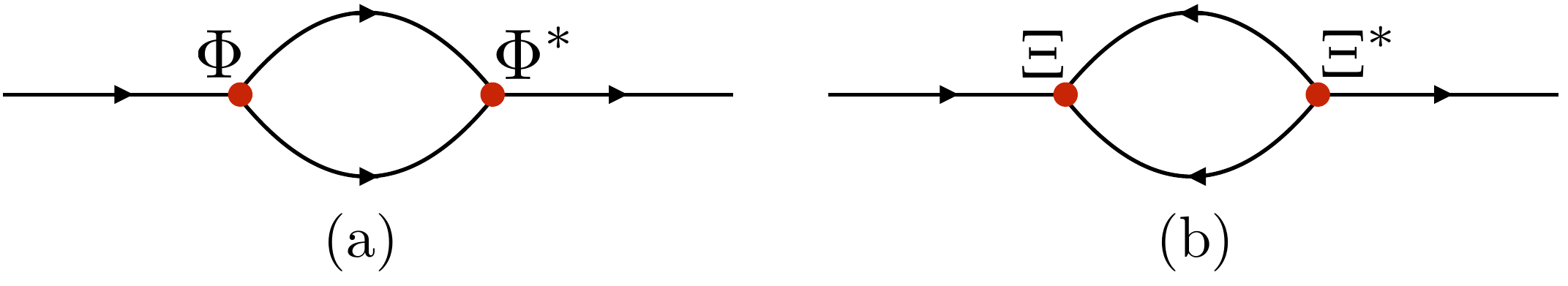}
\caption{Decay (a) and source (b) diagrams.}
\label{s_fig_loops}
\end{figure}

\begin{figure}[b]
\includegraphics[width=\linewidth]{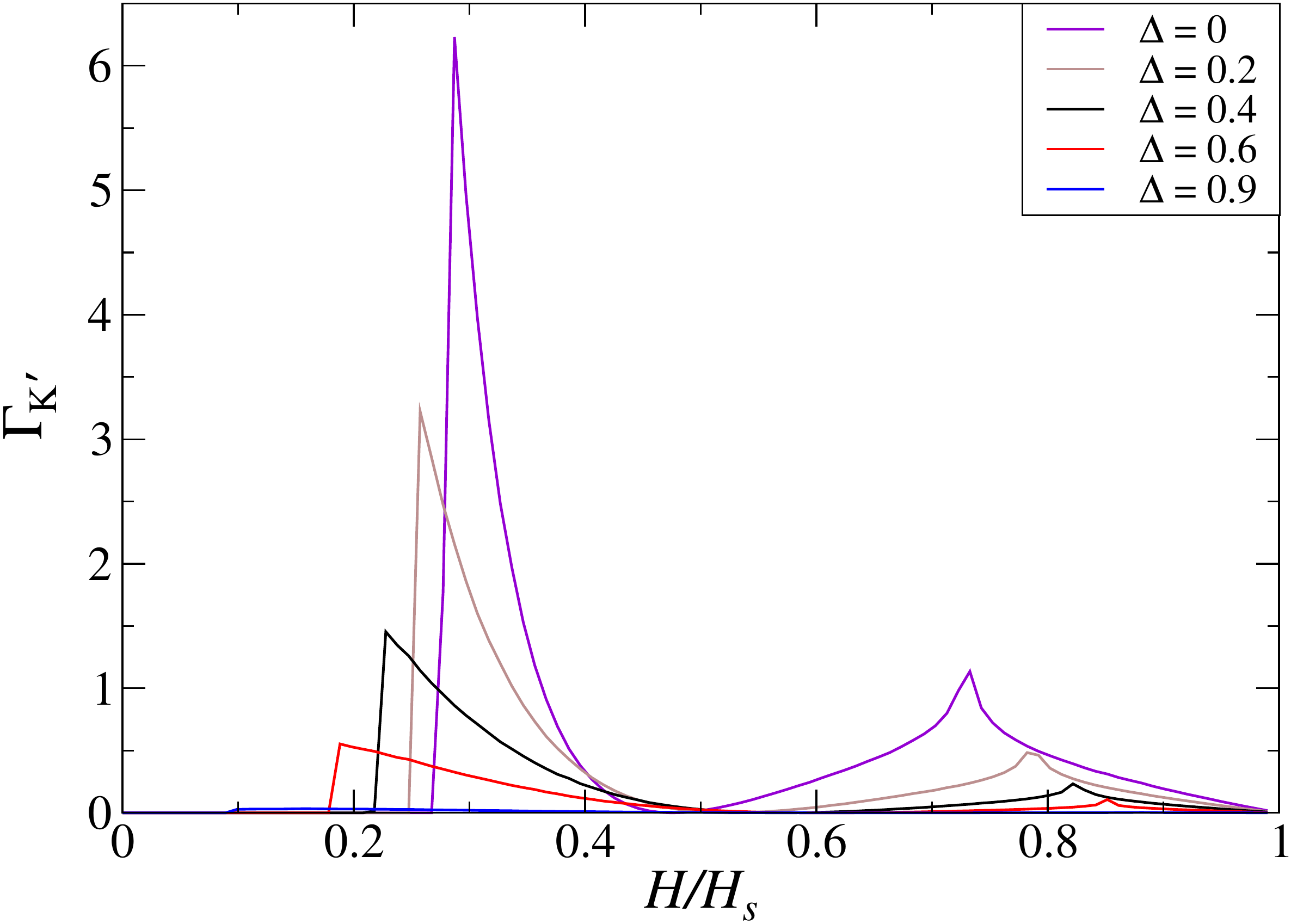}
\vskip -0.3cm
\caption{The on-shell $\Gamma_{\bf k}$ at the K$'$ point vs field for several $\Delta$.
}
\label{s_fig_Gamma_vs_field}
\end{figure}

\subsection{Magnon decays}

Using standard diagrammatic rules, the one-loop decay and source diagrams shown in Fig.~\ref{s_fig_loops} are
\begin{eqnarray}
&&\Sigma_{11}^{(a)}(\mathbf{k},\omega)=\frac{9 J^2 S}{4} 
\sum_{\bf q} \frac{\left| {\Phi}_{{\bf q},{\bf k-q};{\bf k}}\right|^2}
{\omega - \varepsilon_{\bf q} - \varepsilon_{{\bf k}- \bf q}+i\delta}\, ,
\label{s_Sigma_d}\\
&&\Sigma_{11}^{(b)} (\mathbf{k},\omega)=-\frac{9 J^2 S}{4} 
\sum_{\bf q} \frac{\left| {\Xi}_{{\bf q},{\bf -k-q},{\bf k}}\right|^2}
{\omega + \varepsilon_{\bf q} + \varepsilon_{-{\bf k}- \bf q}-i\delta}\, .\quad\quad
\label{s_Sigma_s}
\end{eqnarray}
In our numerical calculations we used $\delta=0.0005J$. 

The on-shell magnon decay rate from ${\rm Im} \Sigma^{(a)}_{11}$ in (\ref{s_Sigma_d})  is 
\begin{eqnarray}
\Gamma_{{\bf k}} = \Gamma_{0}  \sum_{{\bf q}} 
\left| {\Phi}_{{\bf q},{\bf k-q};{\bf k}}\right|^2\delta \left(\omega_{\bf k}-\omega_{\bf q}-\omega_{\bf k-q}\right),
\label{s_eq_gamma}
\end{eqnarray}
with the auxiliary  constant $\Gamma_{0} =3\pi J/4$.

To get a qualitative insight into magnon decays,  one can neglect for a moment the other $1/S$-contributions to the 
spectrum and focus on the effect of the decay part.
Given the field-induced asymmetry of the spectrum, it is clear that the K$'$ point will be prone to decays 
above a threshold field. Our Fig.~\ref{s_fig_Gamma_vs_field} shows the on-shell $\Gamma_{\bf k}$ from 
(\ref{s_eq_gamma}) vs $H$
for this representative ${\bf k}$-point at several values of $\Delta$. One can see  that at large anisotropies, the on-shell
damping reaches unphysically large values.

\begin{figure}[t]
\includegraphics[width=\linewidth]{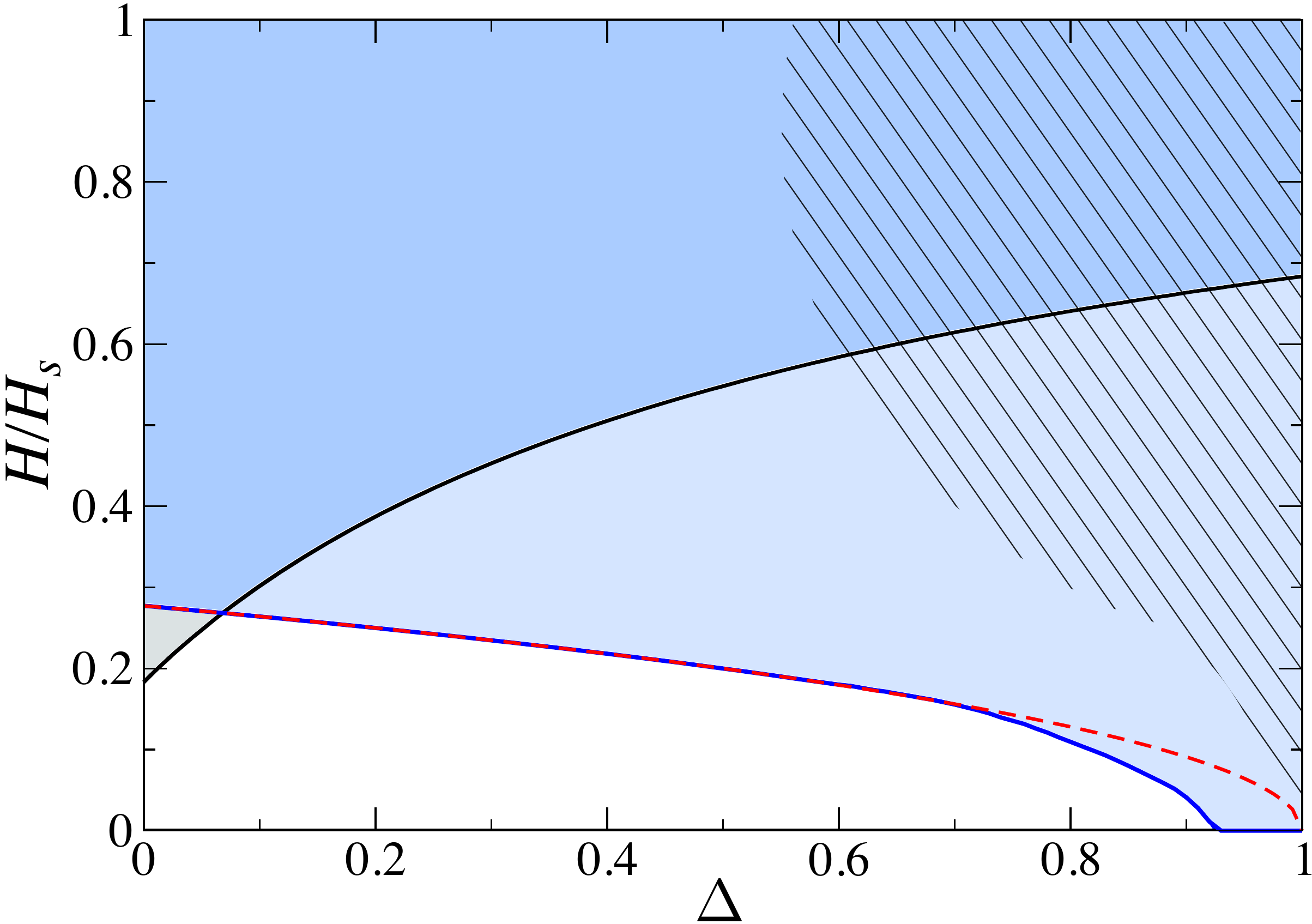}
\caption{The $H\!-\!\Delta$ diagram of the decay thresholds in the umbrella state. 
Color shaded regions represent values of parameters where decays are allowed by kinematic conditions. 
The decay condition $\varepsilon_{\bf K'}=2\varepsilon_{{\bf K}}$ 
 is shown by the dashed line. Decays from a vicinity of the K$'$  point exist above the blue  line. 
 Decays near the $\Gamma$ point, due to the change of curvature of the Goldstone mode, exist 
above the black line. For $\Delta \!>\!0.92$ decays exist in zero field \cite{s_triangle}. While semiclassically
the umbrella state is stable for any $\Delta\!<\!1$, quantum fluctuations lead to proliferation of coplanar states. 
The non-umbrella region for $S\!=\!1/2$ is sketched from Ref.~\cite{s_yamamoto} by a striped pattern.
}
\label{s_fig_diagram}
\vskip -0.2cm
\end{figure}

\emph{Kinematics.}---%
A more systematic approach to the on-shell decays involves a consideration of the general decay conditions \cite{s_RMP},
$\varepsilon_{{\bf k}}=\varepsilon_{{\bf q}}+\varepsilon_{{\bf k-q}}$. The results are presented in the 
$H\!-\!\Delta$ diagram in Fig.~\ref{s_fig_diagram}.
There are two main decay channels here. 
First, the threshold field value for decays in the proximity of K$'$ was found numerically (blue line). 
However, for $\Delta\alt 0.7$ it becomes precisely the threshold condition for the  decays directly from the K$'$ point into
two magnons at the equivalent K points: $\varepsilon_{\text{K}'}=2\varepsilon_{\text{K}}$, 
which is given analytically by $H^{*}=\sqrt{(1-\Delta)/(13-\Delta)}$ (dashed line). Note that this channel is 
permitted by the commensurability of the umbrella state, which retains the 
$3{\bf K}\!=\!0$ property of the 120$\degree$ structure.
The second channel is due to the change of curvature of the spectrum to a negative one  in the vicinity of the $\Gamma$ point 
\cite{s_99,s_pitaevskii}. Expansion of the spectrum to $k^3$ order yields
\begin{equation}
\varepsilon_{|{\bf k}|\rightarrow 0} \simeq \sqrt{\frac{\lambda}{2}} |{\bf k}|-|{\bf k}|^3 
\left[ \frac{1}{16}\sqrt{\frac{\lambda}{2}} \left( \frac{5}{2}-\frac{1}{\lambda}\right) +\frac{h\sqrt{3}}{24}\cos 3\phi\right],
\end{equation}
where $\phi=\tan^{-1} (k_y/k_x)$. The solution of that implicit equation is shown in Fig.~\ref{s_fig_diagram} as a black line.

\begin{figure}[t]
\includegraphics[width=\linewidth]{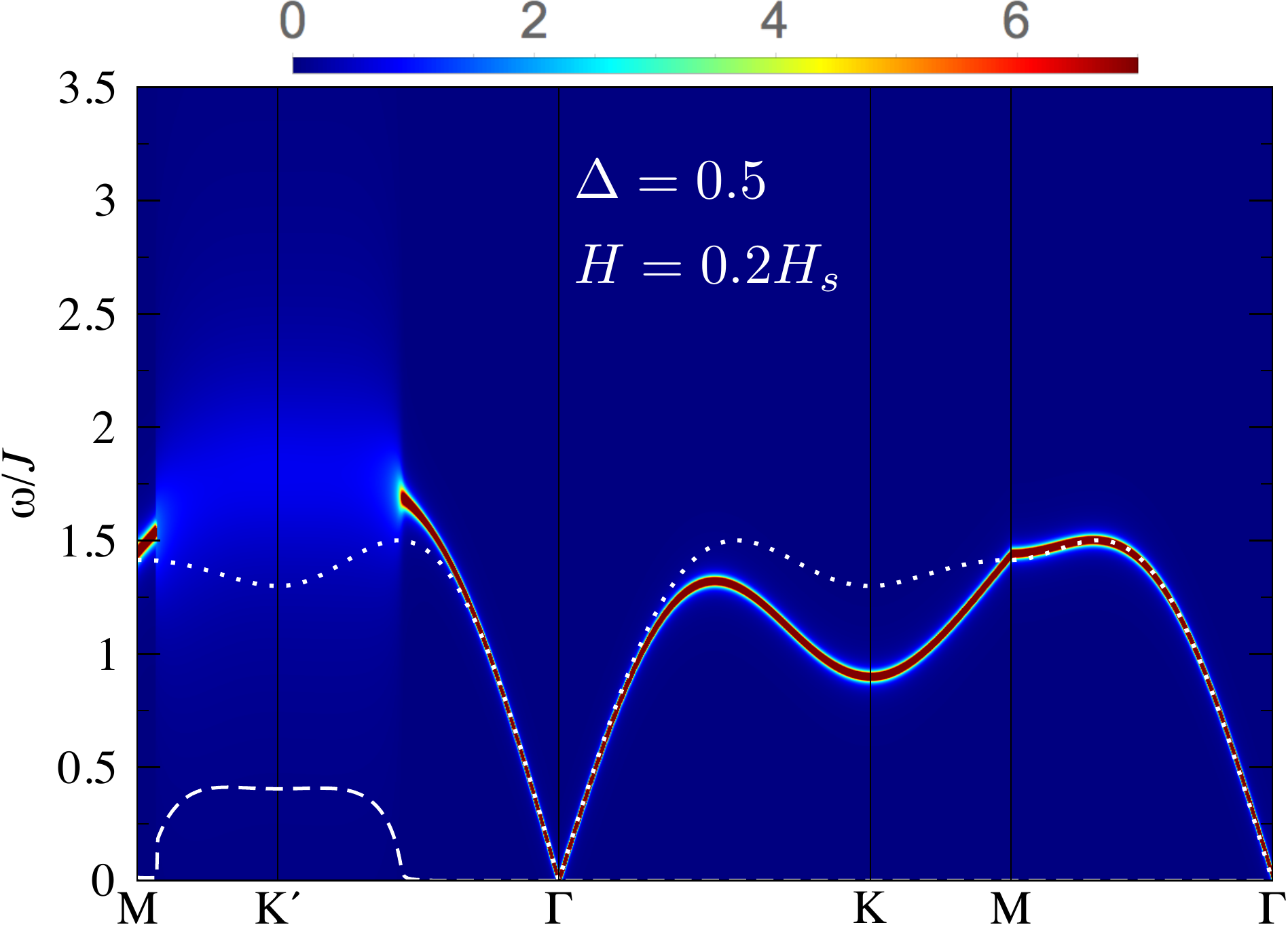}
\caption{
Intensity plot of the magnon spectral function along the MK$'\Gamma$KM$\Gamma$ path  
with the iDE $\Gamma_{\bf k}$, shown by the dashed line. 
$S\!=\!1/2$, $H\!=\!0.2H_s$, and $\Delta\!=\!0.5$. 
Dotted line is the LSW spectrum for $H\!=\!0$.
}
\label{s_fig_iDE}
\vskip -0.2cm
\end{figure}

\emph{Dyson's equation.}---%
In addition to the on-shell damping, one can employ a straightforward self-consistent approach,
which we refer to as iDE, which consists of solving the off-shell Dyson's equation (DE) for $\Gamma_{\bf k}$ of the
magnon pole of the Green's function where only the imaginary part of the the magnon self-energy 
is retained \cite{s_triangle}
\begin{equation}
\Gamma_{\bf k} = -\text{Im}\,\Sigma_{\bf k}\left( \varepsilon_{\bf k}+i\Gamma_{\bf k}\right).
\label{s_iDE}
\end{equation} 
This method accounts for a damping of the decaying initial-state magnon and regularizes   
singularities due to   two-magnon continuum. 
We present some of the results of this approach in Fig.~\ref{s_fig_iDE}, 
which shows the magnon spectral function (intensity map) with the broadening $\Gamma_{\bf k}$ (dashed line)
for representative field and $\Delta$. The self-consistency clearly mitigates the unphysically large
values of the on-shell $\Gamma_{\bf k}$, but its values remain quite significant.
Note that in the iDE self-consistent solution, decays occur at a lower field than suggested by the on-shell consideration.

Altogether, both the on-shell and the iDE  considerations suggest a significant
field-induced broadening of quasiparticle peaks due to magnon decays in modest fields
in a wide vicinity of the K$'$  points.   

\subsection{Hartree-Fock corrections}

For a more consistent consideration of the $\omega$-dependence of the magnon spectral function we also 
need to include all contributions to the one-loop magnon self-energy of the same  $1/S$-order \cite{s_RMP}.
For that we need the source diagram (\ref{s_Sigma_s}) and the Hartree-Fock corrections. There are two contributions
to the latter. First is from the four-magnon interactions (quartic terms) and the second is
from the cubic terms due to quantum corrections to the out-of-plane canting angle  of spins.

Four magnon interaction terms originate from  \eqref{s_eq_Heven} and can be decoupled using Hartree-Fock averages
\begin{eqnarray}
n&=&\langle a^\dagger_i a^{\phantom{\dagger}}_i \rangle =\sum_{\bf k} v_{\bf k}^2,\\
m&=& \langle a^\dagger_i a^{\phantom{\dagger}}_j \rangle =\sum_{\bf k} v_{\bf k}^2 \gamma_{\bf k},\\
\delta &=&\langle a^{\phantom{\dagger}}_i a^{\phantom{\dagger}}_i \rangle =\sum_{\bf k} u_{\bf k} v_{\bf k} ,\\
\bar{\Delta} &=&\langle a^{\phantom{\dagger}}_i a^{\phantom{\dagger}}_j \rangle 
=\sum_{\bf k} u_{\bf k} v_{\bf k}\gamma_{\bf k},
\end{eqnarray}
to obtain correction to the magnon energy spectrum after decoupling and Fourier transformation
\begin{eqnarray} 
&&\delta {\cal{H}}_2^{(4)} =3J\sum_{{\bf k}} 
\left(\delta A^{(4)}_{\bf k} +\delta C^{(4)}_{\bf k}\right) a^\dagger_{{\bf k}} a^{\phantom{\dag}}_{{\bf k}} 
\\
&&\phantom{\hat{\cal H}^{(2)}{(4)}=\quad\quad\quad }
-\frac{\delta B^{(4)}_{\bf k}}{2} \left( a^\dagger_{{\bf k}} a^{\dagger}_{-{\bf k}}
+{\rm H.c.}\right), \nonumber
\end{eqnarray}
where the coefficients are given by
\begin{eqnarray}
\delta A^{(4)}_{\bf k} &= &-d_1- \gamma_{\bf k} d_2 \nonumber\\
d_1&= &\left[ \bar{\Delta} \left(\Delta+\frac12\right) +m \left( \Delta - \frac12 \right) +n \right] \nonumber\\
&&- h^2 \left( \Delta+ \frac12 \right) \left( 2n+ \bar{\Delta} +m \right)\nonumber\\
d_2&=&  \left[ m+n \left( \Delta-\frac{1}{2}\right) +\frac{\delta}{2} \left( \Delta+ \frac{1}{2}\right)\right] \nonumber\\
&&- h^2  \left( \Delta+ \frac12 \right) \left( 2m+ \frac{\delta}{2} +n \right),
\\
\delta B^{(4)}_{\bf k} &=& d_3+\gamma_{\bf k} d_4\nonumber\\
d_3&= & \frac12 \left[ m \left( \Delta+\frac12 \right) +\bar{\Delta} \left( \Delta-\frac12 \right)\right]\nonumber\\
&&- \frac{h^2}{2} \left( \Delta+\frac12 \right) \left( \bar{\Delta}+m \right)\nonumber\\
d_4&=& \left[ \bar{\Delta}+n \left( \Delta +\frac12 \right)  +\frac{\delta}{2} \left( \Delta -\frac12 \right)\right]\nonumber\\
&&- h^2 \left( \Delta+\frac12 \right) \left( n+2\bar{\Delta} +\frac{\delta}{2}\right)
\\
\delta C^{(4)}_{\bf k} &=& - \sqrt{3} n h \bar{\gamma}_{\bf k}
\end{eqnarray}
and we have introduced  a notation $h\! =\!\sin \theta$. Using Bogolyubov transformation \eqref{s_eq_bogolyubov} in 
this Hamiltonian gives
\begin{eqnarray}
\delta {\cal{H}}_2^{(4)} =\sum_{{\bf k}} 
\varepsilon^{(4)}_{\bf k} c^\dagger_{{\bf k}} c^{\phantom{\dag}}_{{\bf k}}
-\frac{V^{od,(4)}_{\bf k}}{2} \left( c^\dagger_{{\bf k}} c^{\dagger}_{-{\bf k}}
+{\rm H.c.}\right),
\label{s_eq_dH4}
\end{eqnarray}
where 
\begin{eqnarray}
&&\varepsilon^{(4)}_{\bf k}=3J\left(\frac{A_{\bf k} \delta A^{(4)}_{\bf k} - B_{\bf k} \delta B^{(4)}_{\bf k}}{\sqrt{A_{\bf k}^2 
- B_{\bf k}^2}} +\delta C^{(4)}_{\bf k}\right),\\
&&V^{od,(4)}_{\bf k} = 
-3J\,\frac{B_{\bf k} \delta A^{(4)}_{\bf k} -A_{\bf k} \delta B^{(4)}_{\bf k}}{\sqrt{A_{\bf k}^2 - B_{\bf k}^2}},
\end{eqnarray}
and $\delta C^{(4)}_{\bf k}$ is not affected by diagonalization as it is odd in ${\bf k}$.
Thus, the Hartree-Fock corrections from the four-magnon interaction to the energy spectrum is 
$\varepsilon^{(4)}_{\bf k}$.

\subsection{Quantum correction from the three-boson terms}

The Hartree-Fock decoupling of the three-magnon term (\ref{s_eq_H3ij}) gives a correction to 
the canting angle $\theta$ via 
\begin{equation}
\hat{\cal H}^{(1)}+\hat{\cal H}^{(3)}_1= \left( V_1+\delta V_1\right)
\sum_i \left( a^{\phantom{\dag}}_{i}+ a^{{\dag}}_{i}\right),
\end{equation}
where $\delta V_1\!=\!3J\sqrt{S/2} (\Delta+1/2) \sin 2\theta (\bar{\Delta}+m+n)$,
adding to the contribution of the  Hamiltonian in (\ref{s_eq_Hodd}), $\hat{\cal H}^{(1)}$, 
which was vanishing due to energy minimization,
with $V_1\!=\!\sqrt{S/2} \cos \theta \left( H-6JS \sin \theta \left( \Delta+1/2 \right) \right)$.
Now, both terms should be (re)balanced to zero:
$\delta \hat{\cal H}^{(3)}_1 + \delta \hat{\cal H}^{(1)}=0$, giving the $1/S$ correction to the canting angle \cite{s_99}
\begin{equation}
\delta \theta = \frac{\sin \theta_0 \left( \bar{\Delta}+m+n \right) }{S \cos \theta_0},
\end{equation}
or, using $1/S\ll 1$,
\begin{equation}
\sin \theta = \sin \theta_0 \left( 1+ \frac{\bar{\Delta}+m+n}{S}\right).
\end{equation}
This  yields a correction to the harmonic term \eqref{s_eq_H2} 
\begin{equation}
\delta \hat{\cal H}^{(3)}_2 = \frac{\partial \hat{\cal H}^{(2)}}{\partial \theta} \delta \theta,
\end{equation}
which, after Fourier transform, can be written as
\begin{align}
\delta \hat{\cal H}^{(3)}_2 &=3JS\sum_{{\bf k}} 
\left(\delta A^{(3)}_{\bf k} +\delta C^{(3)}_{\bf k}\right) 
a^\dagger_{{\bf k}} a^{\phantom{\dag}}_{{\bf k}} \\
&\phantom{\quad\quad\quad\quad\quad}
-\frac{\delta B^{(3)}_{\bf k} }{2}\left( a^\dagger_{{\bf k}} a^{\dagger}_{-{\bf k}}
+{\rm H.c.}\right), \nonumber
\end{align}
with the parameters given by
\begin{eqnarray}
&&\delta A^{(3)}_{\bf k}=-\left(1+\gamma_{{\bf k}}\right) \left( \Delta+\frac{1}{2}\right)\sin 2 \theta\cdot\delta\theta, \\
&&\delta B^{(3)}_{\bf k}=\gamma_{{\bf k}}\left( \Delta+\frac{1}{2}\right)\sin2 \theta\cdot\delta\theta ,\\
&&\delta C^{(3)}_{\bf k}=\sqrt{3} \bar{\gamma}_{{\bf k}}\cos\theta\cdot\delta\theta .
\end{eqnarray}
Similarly to the quartic terms (\ref{s_eq_dH4}), this gives 
\begin{eqnarray}
&&\varepsilon^{(3)}_{\bf k} =
3JS\left( \frac{A_{\bf k} \delta A^{(3)}_{\bf k}   - B_{\bf k} \delta B^{(3)}_{\bf k} }{\sqrt{A_{\bf k}^2 - B_{\bf k}^2}} 
+\delta C^{(3)}_{\bf k}\right),\\
&&V^{od,(3)}_{\bf k} = -3JS\left(\frac{B_{\bf k}\delta A^{(3)}_{\bf k} -A_{\bf k}\delta B^{(3)}_{\bf k}}
{\sqrt{A_{\bf k}^2 - B_{\bf k}^2}}\right).
\end{eqnarray}
Altogether, the Hartree-Fock correction to the magnon energy spectrum is given by
\begin{equation}
\Sigma^\text{HF}(\mathbf{k})=\varepsilon^{(3)}_\mathbf{k}+\varepsilon^{(4)}_\mathbf{k}.
\label{s_HP1}
\end{equation}

\begin{figure*}
\includegraphics[width=0.97\linewidth]{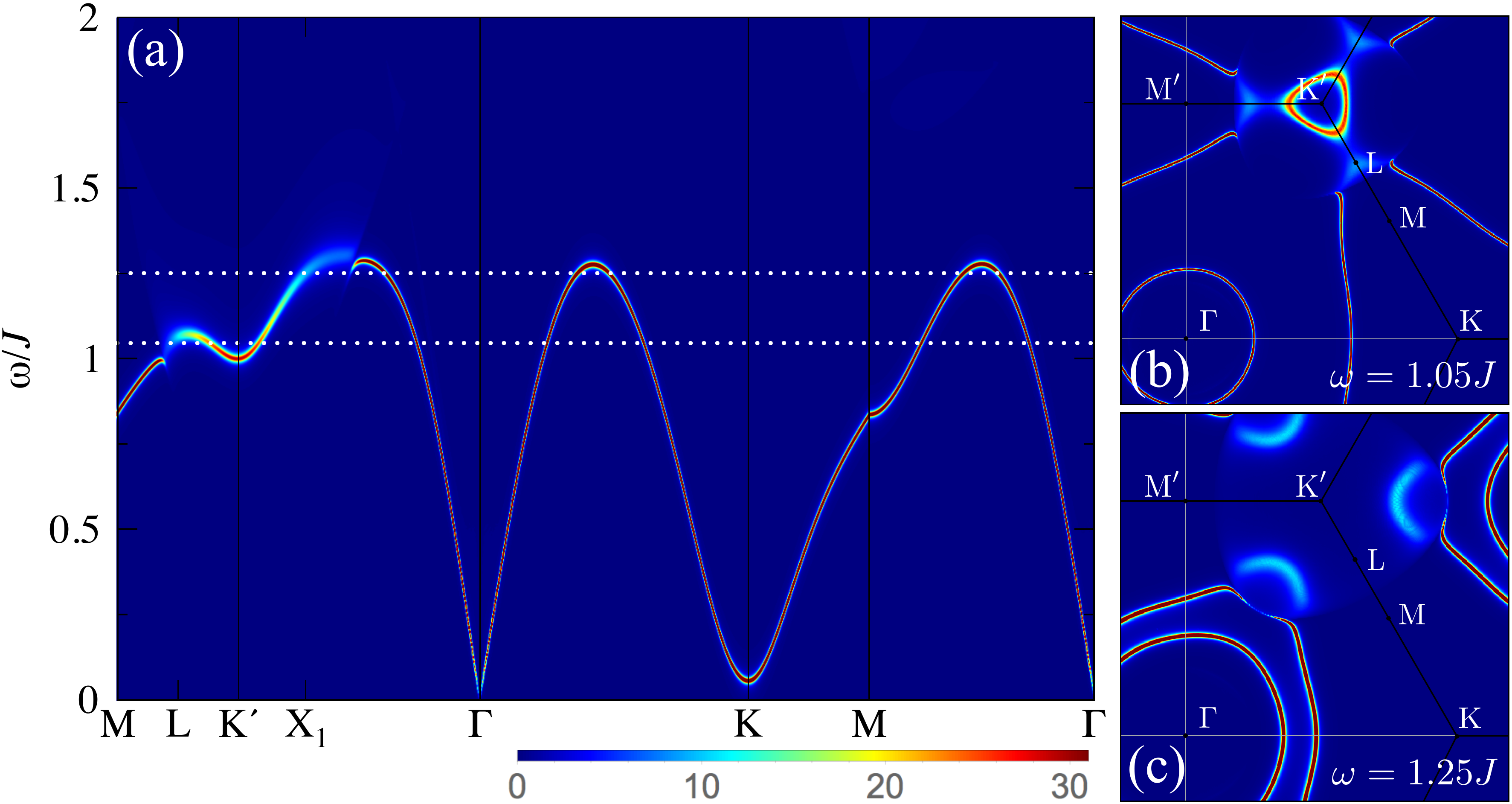}
\caption{(a) The ${\bf k}-\omega$ intensity plot of the spectral function 
$A(\mathbf{k},\omega)$ along the MK$'\Gamma$KM$\Gamma$ path. Dotted lines indicate constant energy cuts in 
(b) and (c), where intensity plots at $\omega=1.05J$ and $\omega=1.25J$ are shown. $S=1/2$, $\Delta=0.9$, $H=0.2H_s$.
}
\label{s_fig_spectral}
\vskip -0.3cm
\end{figure*}

\vspace{-0.6cm}

\subsection{Magnon spectral function}

The  
spectral function, $A(\mathbf{k},\omega)\!=\!-\frac{1}{\pi}\text{Im}G(\mathbf{k},\omega)$, 
with the Green's function  $G(\mathbf{k},\omega)=[\omega-\varepsilon_\mathbf{k}-\Sigma(\mathbf{k},\omega)+i\delta]^{-1}$,  
can now be calculated using the full expression for the $1/S$, one-loop, 
$\omega$-dependent self-energy (\ref{s_Sigma_d}),  (\ref{s_Sigma_s}),  (\ref{s_HP1})
\begin{equation}
\Sigma(\mathbf{k},\omega)=\Sigma^{\text{HF}}({\bf k})+
\Sigma_{11}^{(a)} \left( \mathbf{k},\omega \right)+\Sigma_{11}^{(b)} \left( \mathbf{k},\omega\right).
\end{equation}
In our calculations,  we also kept the source self-energy term on-shell, i.e., 
$\Sigma_{11}^{(b)} (\mathbf{k},\varepsilon_\mathbf{k})$, in order to avoid interaction with negative frequency excitations, 
see \cite{s_triSqw}.

The results of the calculations of the spectral function $A\left(\mathbf{k},\omega \right)$ are shown 
as intensity plots in Fig.~\ref{s_fig_spectral}(a)-(c) for $S=1/2$, $\Delta=0.9$, and $H=0.2H_s$. 
A cut along the high-symmetry path in Fig.~\ref{s_fig_spectral}(a) exhibits  
a strong spectrum renormalization, $\sim 30\%$, and a significant broadening of the quasiparticle peaks 
for an extensive range of momenta in the vicinity of K$'$ corners of the BZ.
The latter is due to an overlap of the one-magnon spectrum with the two-magnon continuum
associated with the  roton-like excitations at the K points. One can also see 
well-pronounced termination points with distinctive bending of spectral lines and other non-Lorentzian features
in Fig.~\ref{s_fig_spectral}(a).
The 2D intensity maps of the constant-energy cuts of $A\left(\mathbf{k},\omega \right)$ at $\omega\!=\!1.05J$ 
and $\omega\!=\!1.25J$  are shown in Fig.~\ref{s_fig_spectral}(b) and (c), where one can see multiple signatures of 
the broadening, spectral weight redistribution, and termination points.

\subsection{Dynamical structure factor}

\begin{figure}[t]
\includegraphics[width=0.7\linewidth]{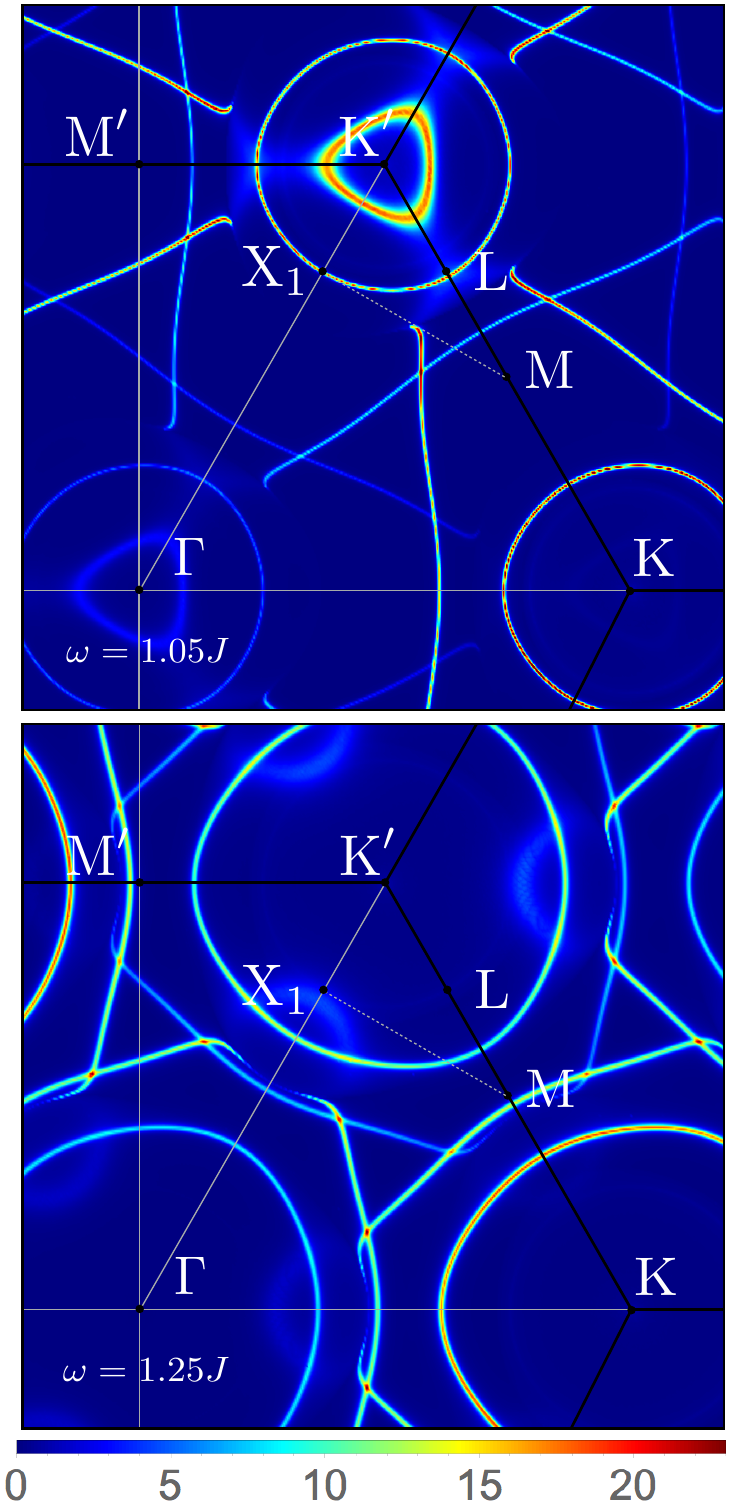}
\caption{Intensity plots of the constant energy cuts  of $\mathcal{S} ({\bf q},\omega)$ 
at the same $\omega$ and parameters as in Fig.~\ref{s_fig_spectral}(b) and (c).}
\label{s_fig_strfac_cut}
\end{figure}

We approximate the dynamical structure factor \cite{s_triSqw} as a sum of the diagonal terms of
\begin{equation}
\mathcal{S}^{\alpha_0 \beta_0} ({\bf q},\omega) =\frac{i}{\pi} \text{Im} 
\int_{-\infty}^{\infty} dt e^{i\omega t}\langle T S^{\alpha_0}_{{\bf q}}(t) S^{\beta_0}_{-{\bf q}}(0) \rangle.
\label{s_eq_strfac}
\end{equation}
The spin rotation to the local reference frame yields (keeping the leading $1/S$ terms)
\begin{eqnarray}
\mathcal{S}^{x_0 x_0}(\mathbf{q},\omega)=
\frac{1}{4}\bigg[\left(\mathcal{S}^{yy}_{\mathbf{q}+}+\mathcal{S}^{y y}_{\mathbf{q}-}\right)
+2i\sin\theta\left(\mathcal{S}^{x y}_{\mathbf{q}+}-\mathcal{S}^{x y}_{\mathbf{q}-}\right)\nonumber\\
+\sin^2 \theta 
\left(\mathcal{S}^{x x}_{\mathbf{q}+}+\mathcal{S}^{x x}_{\mathbf{q}-}\right) 
+\cos^2 \theta \left(\mathcal{S}^{zz}_{\mathbf{q}+}+\mathcal{S}^{z z}_{\mathbf{q}-}\right)
\bigg], \quad 
\label{Sx0x0} 
\end{eqnarray}
\begin{eqnarray}
\mathcal{S}^{y_0 y_0}(\mathbf{q},\omega)=\mathcal{S}^{x_0 x_0}(\mathbf{q},\omega),
\quad\quad\quad\quad\quad\quad\quad\quad\quad\quad 
\label{Sy0y0}\\
\mathcal{S}^{z_0 z_0}(\mathbf{q},\omega)=
\cos^2 \theta\mathcal{S}^{x x}_{\mathbf{q}}+\sin^2\theta\mathcal{S}^{z z}_{\mathbf{q}},
\quad\quad\quad\quad\quad\quad 
\end{eqnarray}
where we used  $\mathcal{S}^{x y}_{\mathbf{q}}=-\mathcal{S}^{yx}_{\mathbf{q}}$
and the shorthand notations from Ref.~\cite{s_triSqw},
$\mathcal{S}_{\mathbf{q}\pm}\equiv \mathcal{S}\left(\mathbf{q}\pm\mathbf{Q},\omega\right)$.

We note that the off-diagonal component, $\mathcal{S}^{x_0 y_0} ({\bf q},\omega)$, 
is also non-zero, but its contribution to the total structure factor for an unpolarized 
beam is canceled exactly by its partner, $\mathcal{S}^{y_0 x_0} ({\bf q},\omega)$, leaving 
the $\mathcal{S}^{x y}_{\mathbf{q}\pm}$ contributions in (\ref{Sx0x0}) and (\ref{Sy0y0}) 
to be the sole unorthodox terms due to the field-induced chiral structure.

Transverse components of $\mathcal{S}({\bf q},\omega)$ are given by
\begin{eqnarray}
&&\mathcal{S}^{xx(yy)}_\mathbf{q}=\frac{S}{2}\Lambda_{\pm}^2 
\left(u_{\mathbf q} \pm v_{\mathbf q}\right)^2 A(\mathbf{q},\omega),\\
&&\mathcal{S}^{x y}_{\mathbf{q}}=i\frac{S}{2}\Lambda_+ \Lambda_- 
\, A(\mathbf{q},\omega),
\end{eqnarray}
where $\Lambda_\pm=1-(2n\pm\delta)/4S$,
and the longitudinal one is
\begin{equation}
\mathcal{S}^{zz}_\mathbf{q}
=\frac{1}{2}\sum_{\bf k} \left( u_{\bf k}v_{\mathbf{q}-\mathbf{k}}+v_{\bf k}u_{\mathbf{q}-
\mathbf{k}}\right)^2 \delta \left( \omega - \varepsilon_{\mathbf{k}}-\varepsilon_{\mathbf{q}-\mathbf{k}}\right).
\end{equation}
Clearly, $\mathcal{S}({\bf q},\omega)$ should feature three
overlapping single-magnon spectral functions,  $A({\bf q},\omega)$ and $A({\bf q}\pm {\bf Q},\omega)$, 
with different kinematic formfactors.

In Fig.~\ref{s_fig_strfac_cut} we present constant energy cuts of $\mathcal{S}(\mathbf{q},\omega)$ 
at the same energies and parameters as in Fig.~\ref{s_fig_spectral}(b) and (c) to demonstrate proliferation of the 
unusual  features discussed for the spectral function.




\begin{thebibliography}{99}
\bibitem{footnote1} The phase diagrams in Refs. \cite{yamamoto} and \cite{eggert} are slightly different.
\end{thebibliography}
\end{document}